\documentclass[letter,12pt]{article}
\usepackage[top=1in,bottom=1in,left=1in,right=1in]{geometry}
\usepackage[T1]{fontenc} 
\usepackage{slashed}
\usepackage{epsfig}
\usepackage{setspace}
\usepackage{comment}
\usepackage{cancel}
\usepackage[top=1in,bottom=1in,left=1in,right=1in]{geometry}

\pdfoutput=1

\usepackage{amsmath,amssymb,amsfonts}
\usepackage{amsthm}         
\usepackage{latexsym}
\usepackage[scr=boondox]{mathalfa}
\usepackage{dsfont}
\usepackage{physics}
\usepackage[normalem]{ulem}
\usepackage{slashed}
\usepackage{simpler-wick}
\usepackage{cancel}
\usepackage{yhmath}
\usepackage{mathdots}
\usepackage{cite}

\usepackage{pgfplots}
\pgfplotsset{compat=1.17}
\usepackage{tikz}
\usepackage{tikz-3dplot}
\usetikzlibrary{
  3d,
  calc,
  fadings,
  patterns,
  decorations.pathreplacing,
  shadows.blur,
  shapes
}

\usepackage{booktabs}
\usepackage{multirow}
\usepackage{tabularx}
\usepackage{array}

\usepackage{gensymb}
\usepackage{siunitx}
\usepackage{pifont}

\usepackage[utf8]{inputenc}
\usepackage{verbatim}

\usepackage{hyperref}

\usepackage{caption}

\numberwithin{equation}{section}

\newcommand{\be}{\begin{equation}}
\newcommand{\ee}{\end{equation}}
\newcommand{\bs}{\begin{subequations}}
\newcommand{\es}{\end{subequations}}

\def\e{{\epsilon}}

\def\s{{\sigma}}

\def\a{{\alpha}}
\def\b{{\beta}}

\def\G{{\Gamma}}
\def\Th{{\Theta}}
\def\d{{\delta}}
\def\z{{\zeta}}
\def\ka{{\kappa}}

\def\D{\Delta}
\def\O{\Omega}

\def\w{\omega}

\def\CB{{\mathcal B}}
\def\CC{{\mathcal C}}
\def\CD{{\mathcal D}}

\def\CF{{\mathcal F}}

\def\CH{{\mathcal H}}

\def\CL{{\mathcal L}}

\def\CO{{\mathcal O}}

\def\CS{{\mathcal S}}

\def\CZ{{\mathcal Z}}

\def\SI{{\mathscr I}}
\def\SJ{{\mathscr J}}

\def\la{\langle}
\def\ra{\rangle}
\newcommand{\dt}{{\text d}}

\def\+{{(+)}}
\def\-{{(-)}}
\def\0{{(0)}}
\def\1{{(1)}}
\def\2{{(2)}}
\def\3{{(3)}}

\def\p{{\partial}}

\def\grav{{\text{grav}}}

\def\eff{{\text{eff}}}
\def\tu{{\tilde{u}}}
\def\tv{{\tilde{v}}}
\def\tr{{\tilde{r}}}
\def\tt{{\tilde{t}}}
\def\tf{{\tilde{f}}}
\def\bu{{\bar{u}}}
\def\bv{{\bar{v}}}

\def\mid{{\text{mid}}}
\def\OS{{\text{OS}}}

\def\TBHPhi{\zeta}
\def\mid{{\text{mid}}}
\def\reg{{\text{reg}}}
\def\grav{{\text{grav}}}

\begin{document}
\begin{titlepage}
\unitlength = 1mm
\hfill CALT-TH 2025-026
\ \\
\vskip 2cm
\begin{center}

\openup 0.5em

{\LARGE{Thermodynamics of a Spherically Symmetric Causal Diamond in Minkowski Spacetime}}

\vspace{0.8cm}
Kwinten Fransen,$^1$ Temple He,$^1$ Kathryn M. Zurek$^1$

\vspace{1cm}

{\it  $^1$Walter Burke Institute for Theoretical Physics \\ California Institute of Technology, Pasadena, CA 91125 USA}\\

\vspace{0.8cm}

\begin{abstract}

We compute a gravitational on-shell action of a finite, spherically symmetric causal diamond in $(d+2)$-dimensional Minkowski spacetime, finding it is proportional to the area of the bifurcate horizon $A_\CB$.  We then identify the on-shell action with the saddle point of the Euclidean gravitational path integral, which is naturally interpreted as a partition function. This partition function is thermal with respect to a modular Hamiltonian $K$. 
Consequently, we determine, from the on-shell action using standard thermodynamic identities, both the mean and variance of the modular Hamiltonian, finding $\langle K \rangle = \langle (\Delta K)^2 \rangle = \frac{A_\CB}{4 G_N}$. 
Finally, we show that modular fluctuations give rise to fluctuations in the geometry, and compute the associated phase shift of massless particles traversing the diamond under such fluctuations.
\end{abstract}

\vspace{1.0cm}
\end{center}
\end{titlepage}
\pagestyle{empty}
\pagestyle{plain}
\pagenumbering{arabic}

\tableofcontents

	%
	%

\singlespacing
	
\section{Introduction}
    
Black hole horizons have been studied extensively in quantum gravity, and from them we have learned much about their nature, including thermodynamical properties \cite{Bekenstein:1973ur,Hawking:1975vcx,Gibbons:1976ue,Witten:2024upt}, microstate interpretation of the entropy~\cite{Bekenstein:1975tw,Zurek:1985gd}, and information theoretic properties \cite{Ryu:2006bv,Ryu:2006ef,Hayden:2007cs,Almheiri:2012rt, Harlow:2014yka,Penington:2019npb}. On the other hand, much less is known about other types of horizons, such as causal horizons created by light rays and cosmological horizons.  Results in recent years suggest that light-sheet horizons share many properties in common with black hole horizons, including certain thermodynamic properties \cite{Laflamme:1987ec,Ryu:2006bv,Casini:2011kv, Hung:2011nu,Perlmutter:2013gua,Jacobson:2015hqa,Verlinde:2019ade,Jacobson:2022gmo, Ciambelli:2019lap, Chandrasekaran:2021hxc, Freidel:2022vjq, Ciambelli:2023mir, Ciambelli:2024swv, Chandrasekaran:2023vzb, Bub:2024nan}.  It may be counterintuitive to understand why a causal diamond created by light rays might be expected to have thermodynamic properties. Some of the known features of the thermodynamics of causal diamonds are given in \cite{Casini:2011kv}, which we will now quickly review, as these ideas are closely related to the analysis we perform in this paper. 

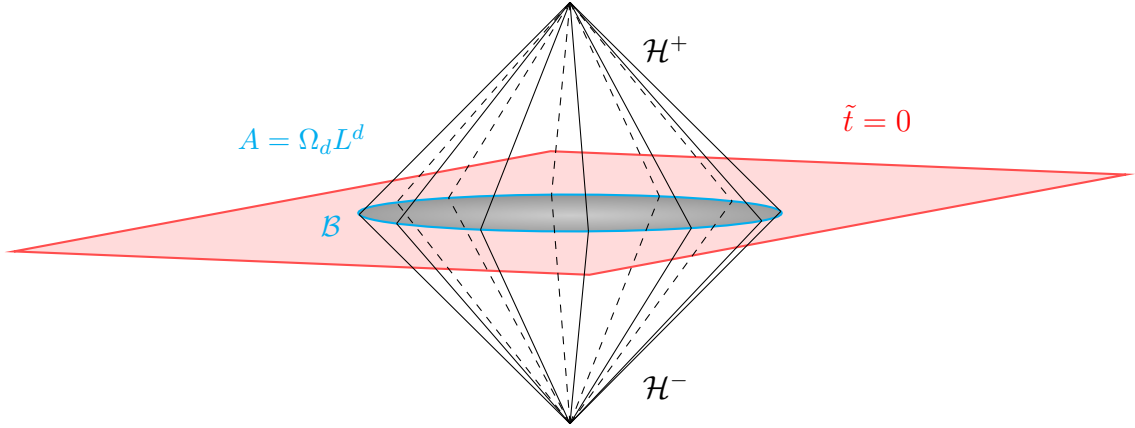
\begin{figure}[t]
\centering
\tdplotsetmaincoords{85}{115} 
\begin{tikzpicture}[tdplot_main_coords, scale=2.8]

  \def\a{3} 
  \def\b{1.5} 
  
  \draw[thick,red,fill=red!20,opacity=0.7]
    (-\a,-\b,0) -- (\a,-\b,0) -- (\a,\b,0) -- (-\a,\b,0) -- cycle;
    \node at (3,3,0.8) {\textcolor{red}{$\tt=0$}};

  \def\R{1} 
  \def\H{1} 
  
  \shade[cyan , thick, shading=radial, inner color=gray!40, outer color=gray!80] (0,0,0) circle (\R);
  \draw[thick,cyan] (0,0,0) circle (\R);
  
  \foreach \angle in {0,30,60,90,120,-30,-60} {
    \draw[] 
      (0,0,\H) -- ({\R*cos(\angle)}, {\R*sin(\angle)}, 0);
    \draw[]
      (0,0,-\H) -- ({\R*cos(\angle)}, {\R*sin(\angle)}, 0);
  }   
  \foreach \angle in {155,180,210,240,260} {
    \draw[dashed] 
      (0,0,\H) -- ({\R*cos(\angle)}, {\R*sin(\angle)}, 0);
    \draw[dashed]
      (0,0,-\H) -- ({\R*cos(\angle)}, {\R*sin(\angle)}, 0);
  }

  \node at (0,-1.4,0.3) {\small\textcolor{cyan}{$A = \O_d L^d$}};
  \node at (0.1,-1.2,-0.1) {\small\textcolor{cyan}{$\CB$}};
  \node at (0,0.5,0.8) { $\CH^+$ };
  \node at (0,0.5,-0.8) { $\CH^-$ };
  
\end{tikzpicture}
\caption{A causal diamond with radius $L$, and area given by $A \equiv \O_dL^d$, where $\O_d$ is the area of a unit $d$-sphere. The complement region to the causal diamond is shaded red and is traced out when obtaining the reduced density matrix $\hat\rho$ associated to the causal diamond.} 
\label{fig:CD}
\end{figure}

When restricting ourselves to a quantum field theory (QFT) inside a causal diamond, we integrate out the degrees of freedom living in the complement region, leaving us a reduced density matrix $\hat\rho$ that describes the degrees of freedom inside the causal diamond (see Figure~\ref{fig:CD}). The entanglement across the bifurcation surface ${\cal B}$ of the causal diamond is then simply the von Neumann entropy $S = -\Tr(\hat\rho \log \hat\rho)$. Furthermore, because $\hat\rho$ is positive semidefinite and Hermitian, we can write
  \begin{equation}
  	\hat\rho = e^{-K},
  \end{equation}
where $K$ is a Hermitian operator known as the modular Hamiltonian.  The modular Hamiltonian generates a symmetry of the system via the unitary $U(\tau) = \hat\rho^{i s} = e^{-i K s}$, which one can easily check using the fact $\Tr(\hat\rho U(s) {\cal O} U(-s)) = \mbox{tr}(\hat\rho{\cal O})$ for any operator ${\cal O}$ localized inside the causal diamond.  One can further check that the correlators involving ${\cal O}$ (under transformations extended to complex time $\tau$) obey the Kubo-Martin-Schwinger periodicity in imaginary time.  
  
Taken together, these facts suggest a thermal behavior for the density matrix $\hat\rho$.  In fact, if we restrict ourselves to the Rindler wedge of the Minkowski half-space $x^1 > 0$, the Bisognano-Wichmann theorem states that the modular Hamiltonian is precisely the generator of boosts along the $x^1$ direction \cite{Bisognano:1975ih,Bisognano:1976za}. In this case, the modular flow is given by
  \begin{equation}
	  x^{\pm}(s) = x^\pm e^{\pm 2 \pi s} = z e^{\pm \frac{\tau}{L}},
  \end{equation}
where $x^{\pm} = x^1 \pm x^0$ are lightcone coordinates, $(\tau,z)$ are Rindler coordinates, and $L$ is a length scale that can be associated to a Rindler observer with some given temperature. Thus, in Rindler coordinates, we see that modular flow corresponds to taking $\tau \rightarrow \tau + 2 \pi L s$. Moreover, the state living in the Rindler wedge is thermal with respect to Rindler time $\tau$, with its density matrix given by
  \begin{equation}
  \hat\rho_\b = \frac{1}{\CZ_\b}e^{-\beta K_\tau} , \qquad \CZ_\b = \Tr(e^{-\b K_\tau}) ,
  \label{eq:thermaldensity}
  \end{equation}
where $\beta = 2 \pi L$ is the inverse temperature and $K = \beta K_\tau + \log \CZ_\b$. Notice that $\CZ_\b$ is the partition function associated to the boost Hamiltonian $K_\tau$. It is necessary to ensure $\Tr\,\hat\rho = 1$. 

The thermal partition function $\CZ_\b$ is the generating function that allows us to compute thermodynamic quantities, and it is related to the free energy $F_\b$ via
  \begin{equation}\label{eqn:Fn}
  F_\beta = - \frac{1}{\beta} \log \CZ_\b.
  \end{equation}
Standard thermodynamic identities then allow us to compute the entropy and its fluctuations using the free energy $F_\b$ (or equivalently $\CZ_\b$), and are given by 
\begin{align}  \label{eq:thermK}
\begin{split}
  \langle K \rangle &= -(1 - \beta \partial_\beta  ) (\beta F_\beta) = S  , \qquad \langle (\Delta K)^2 \rangle = -\beta^2 \partial^2_\beta (\beta F_\beta) .
\end{split}
\end{align}  
This thermodynamic partition function has been shown to describe both stationary black holes and causal diamonds in AdS/CFT \cite{Casini:2011kv, Perlmutter:2013gua, Jacobson:2018ahi, Verlinde:2019ade, Jacobson:2022jir}.
  
We would now like to restrict our analysis to semiclassical spherically symmetric gravitational fluctuations of a large (but finite) causal diamond in $(d+2)$-dimensional Minkowski spacetime. Motivated by \eqref{eq:thermaldensity}, we can equate the partition function -- or equivalently, the gravitational path integral of \cite{Gibbons:1976ue} -- to the trace of the unnormalized reduced density matrix $\rho$, yielding
\begin{align}\label{Z-path}
\begin{split}
	\CZ =  \int [\CD g] e^{-I_\grav[g]} = \Tr\,\rho, 
\end{split}
\end{align}
where $I_\grav[g]$ is the Euclidean action.\footnote{There are subtleties surrounding \eqref{Z-path} in a gravitational context, e.g., see \cite{Harlow:2018tqv, Balasubramanian:2025hns} for recent discussions. } By performing the semiclassical approximation, we are effectively evaluating the path integral at a saddle point and replacing the action with the Euclidean on-shell action $I^\reg_\OS$, so that  
\begin{align}\label{semi}
\begin{split}
	\int [\CD g] e^{-I_\grav[g]} \approx e^{-I_\OS^\reg} \quad\implies\quad \log\CZ \approx - I^{\reg}_\OS.
\end{split}
\end{align}
One of our main results is the computation of this on-shell action associated to the causal diamond, which we determine in \eqref{eqn:Son-shellfixed} to be 
\begin{align}\label{onshell-intro}
\begin{split}
	I_\OS = \frac{A_\CB}{8\pi G_N} ( \ka u_\CB + \a),
\end{split}
\end{align}
where $A_\CB$ is the area of the bifurcate horizon of the unperturbed causal diamond, $u_\CB$ a parameter that formally diverges, $\ka$ a spacetime constant known as the inaffinity and can be identified with the temperature of a particular Rindler trajectory,\footnote{To be precise, $\ka$ is the inaffinity associated to the Gaussian null time $u$, and is defined on any Rindler trajectory to be $\ell^\mu \nabla_\mu\ell^\nu = \ka \ell^\nu$ for $\ell^\mu = \p_u$, with the equality understood to hold when projected onto the trajectory. On the causal horizon, $\ka$ can be identified with the surface gravity (though for non-Killing horizons the surface gravity, as well as the inaffinity, depends on the clock choice). \label{fn:inaffinity}} and $\a$ a spacetime integration constant. Upon regularizing this action, we obtain an on-shell action that is proportional to the area.

To compute the mean and variance of the modular Hamiltonian given the normalized density matrix $\hat\rho = \frac{1}{\CZ} \rho$ analogous to \eqref{eq:thermK}, we utilize the replica method \cite{Callan:1994py, Holzhey:1994we, Calabrese:2004eu, Lewkowycz:2013nqa, Shekar:2024edg}. We construct the $n$-fold replica manifold associated to the causal diamond, which corresponds to gluing $n$ copies of the density matrix with appropriate boundary conditions along the open cuts where the gluing occurs. This gives rise to the generating function
\begin{align}\label{Z-path3-intro}
\begin{split}
	\CZ[n] \equiv \Tr\,\rho^n, \qquad \log \CZ[n] = - I_\OS^\reg[n],
\end{split}
\end{align}
where $I_\OS^\reg[n]$ is the regularized Euclidean on-shell action associated to the replica manifold. Analogous to \eqref{eq:thermK}, we can now use the replica trick to take the $n \to 1$ limit to compute the mean and the variance of the modular Hamiltonian in the original causal diamond, which are given by
\begin{align}
\begin{split}
	\la K \ra = -\lim_{n \to 1} (\p_n - 1) \log \CZ[n], \qquad \la (\D K)^2 \ra =  \lim_{n \to 1} \p_n^2 \log\CZ[n] .
\end{split}
\end{align}
When we evaluate these quantities using the on-shell action associated to the $n$-fold replica manifold, we obtain in \eqref{eqn:Sform} and \eqref{eqn:Sform2}
\begin{align}\label{eqn:intro:KdK}
\begin{split}
	\la K \ra = \la (\D K)^2 \ra = \frac{A_\CB}{4 G_N} .
\end{split}
\end{align}
This result is consistent with previous calculations in the context of AdS/CFT \cite{Perlmutter:2013gua,DeBoer:2018kvc, Verlinde:2019ade}, in flat spacetime \cite{Banks:2021jwj,Verlinde:2022hhs, Gukov:2022oed, He:2024vlp},\footnote{A similar result was also obtained for area fluctuations along a stretched horizon in \cite{Ciambelli:2025flo}.} and also in cosmological spacetimes \cite{Aalsma:2025bcg}.

Furthermore, it is clear from above that the variable conjugate to the modular Hamiltonian $K$ is the replica index $n$. As such, it is also natural to study the fluctuations of $n$ for any fixed modular Hamiltonian by implementing a Legendre transform. This question to our knowledge has not been explored in the literature, and we discover a simple result relating the variance of $K$ to that of $n$, which is given in \eqref{Dn-2} to be
\begin{align}
	\la (\D n)^2 \ra = \frac{1}{\la (\D K)^2 \ra} .
\end{align}
This result is analogous to the more familiar example of a thermal system, where we can study the relation between energy and temperature fluctuations.

\begin{figure}[t]
\centering
\begin{tikzpicture}
    \draw[->] (0,-4) -- (0,4) node[above] {$t$};
    \draw[->] (-1,0) -- (4,0) node[right] {$r$};

    \draw[thick, black] (0,3) -- (3,0) -- (0,-3) --  cycle;
    \filldraw [cyan] (3,0) circle (2pt);
    \node at (3.3,0.3) {\small\textcolor{cyan}{$\CB$}};
    \node at (1,2.8) { $\CH^+$ };
    \node at (1,-2.8) { $\CH^-$ };
    \node at (1,-1) {\color{red} $\CH_s$ };

    \draw[thick, red, domain=1:3, samples=100, smooth] 
        plot (-{\x} + 3, {sqrt(\x^2 - 1)});
    \draw[thick, red, domain=1:3, samples=100, smooth] 
        plot (-{\x} + 3, {-sqrt(\x^2 - 1)});
        
    \draw[->, rotate around = {33:(1.71,0.81)}, thick] (1.71,0.81) -- (1.71,1.21) node[left] {\tiny $u$};
    \draw[->, rotate around = {-135:(1.71,0.81)}, thick] (1.71,0.81) -- (2.11,0.81) node[left] {\tiny $r$};
    
     \draw[thick] (2,-3) -- (3,-3);
    \draw[thick] (2,-2.9) -- (2,-3.1); 
    \draw[dashed] (2,-2.9) -- (2,0);
    \draw[thick] (3,-2.9) -- (3,-3.1);   
    \draw[dashed] (3,-2.9) -- (3,0);
    \node at (2.5,-3.7) { $s_\CB = 2L \sqrt{\frac{|\D K|}{K d}}$ };
    
     \draw[thick] (0,3.5) -- (3,3.5);
    \draw[thick] (0,3.4) -- (0,3.6); 
    \draw[dashed] (0,3.4) -- (0,3);
    \draw[thick] (3,3.4) -- (3,3.6);   
    \draw[dashed] (3,3.4) -- (3,0);
    \node at (1.5, 3.8) { $L$ };
    
     \draw[thick] (-0.1,3) -- (-0.1,2.8);
    \draw[thick] (-0.2,3) -- (0,3); 
    \draw[thick] (-0.2,2.8) -- (0,2.8);   
    \node at (-0.5, 2.9) { $s_\CC$ };
    
    \filldraw [black] (2,0) circle (2pt);
    \node at (2.4, 0.2) {\tiny{$u_\mid$}};
\end{tikzpicture}
\caption{A spacetime diagram of a causal diamond of size $L$ in $d+2$ dimensions with $d$ angular directions suppressed. Under modular fluctuations, the causal horizons of the diamond can become deformed to a stretched horizon $\CH_s$, with its midpoint labeled to be at $u=u_\mid$. The separation between $\CH_s$ and the bifurcate horizon $\CB$ is given by $s_\CB = 2 L \sqrt{\frac{|\D K|}{K d}}$, where $\D K$ is the fluctuation of the modular Hamiltonian. } 
\label{fig:SH}
\end{figure}
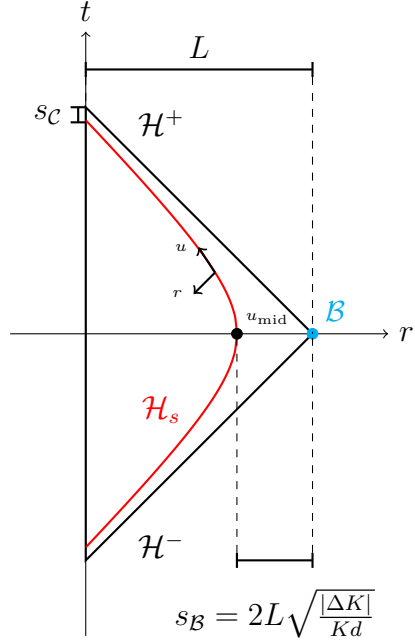

Ultimately, we would like to compute geometric observables of the causal diamond. Because the on-shell action, being computed from the Einstein-Hilbert action, is a geometric object, this suggests fluctuations of the density matrix computed from this action, namely $\la (\D K)^2 \ra$, are physical fluctuations in the geometry. In particular, we find that fluctuations in $K$ manifest in the geometry as shifts in the location of the causal horizon, given in \eqref{eqn:dKtophi1} to be (for $\ka = \frac{1}{2L}$) 
\begin{align}
	s_\CB =  2 L \sqrt{\frac{|\D K|}{K d} },
\end{align}
where $L$ is the radius of the causal diamond (see Figure~\ref{fig:CD}). An identical scaling was determined in flat~\cite{Verlinde:2019xfb} and Anti-de Sitter (AdS) \cite{Verlinde:2019ade} spacetimes, and a similar effect was derived for JT gravity~\cite{Gukov:2022oed}. In effect, the causal horizon becomes perturbed into a Rindler, or stretched horizon, and the corner on which the bifurcate horizon is located becomes smoothed (see Figure~\ref{fig:SH}).

To fully connect our analysis to a gauge-invariant ``observable'' in a thought experiment, we compute the difference in phase between a photon traversing the unperturbed horizon and one traversing the perturbed geometry corresponding to a stretched horizon. Given that the photon has frequency $\w_0$, we find that this difference in phase is to leading order given in \eqref{phase-shift-fin} and \eqref{phase-shift-fin2} to be 
\begin{align}\label{phase-shift-intro}
\begin{split}
	\D\psi = 2\w_0 s_\CB = 4L\w_0 \sqrt{\frac{|\D K|}{Kd} } .
\end{split}
\end{align}
This is precisely the phase difference arising from the Shapiro time delay, which we compute in Appendix~\ref{app:photon}, and importantly only depends on physical quantities measured in the laboratory frame, namely $\w_0$ and $s_\CB$.

The outline of the paper is as follows. In Section~\ref{prelim}, we begin by reviewing the setup of a spherically symmetric causal diamond in Minkowski spacetime, introducing our conventions and coordinate systems.  In Section~\ref{sec:onshell}, we compute an on-shell action associated to the causal diamond. We then in Section~\ref{sec:thermo} identify the Euclidean path integral via the saddle point approximation with the on-shell action and derive the mean and variance of the modular Hamiltonian via thermodynamic identities. The geometric implications of these fluctuations are explored in Section~\ref{geometry}. Finally, in Section~\ref{discussion}, we conclude and discuss some future directions.

\section{Preliminaries}\label{prelim}

Consider the line element describing $(d+2)$-dimensional Minkowski spacetime, which is given in outgoing null coordinates by
\begin{align}\label{eqn:minkowski}
\begin{split}
	\dt s^2 = - \dt \tu^2 - 2\,\dt \tu\,\dt \tr + \tr^2\,\dt \O_d^2,
\end{split}
\end{align}
where $\tu = \tt-\tr$ is the outgoing null time, and $\dt\O_d^2$ is the line element of the $d$-dimensional transverse unit sphere. A causal diamond with length $L$ centered at the origin\footnote{In \cite{Bub:2024nan}, the origin corresponds to the top tip of the causal diamond. However, it is more convenient for us in this paper to choose the more symmetric setup where the origin is the center of the causal diamond.} corresponds to choosing the coordinate $\tu$ to fall in the range $-L \leq \tu \leq L - 2\tr$. It was shown in \cite{Ciambelli:2025flo} that after performing the coordinate transform
\begin{align}\label{diffeo-def}
\begin{split}
	\tu = - L + \frac{1}{\ka} e^{\ka u + \a}, \qquad \tr = L - \frac{1}{2\ka} e^{\ka u + \a} - r e^{-\ka u - \a},
\end{split}
\end{align}
we can rewrite the line element \eqref{eqn:minkowski} in Gaussian null form, namely
\begin{align}\label{eqn:metric}
\begin{split}
	\dt s^2 &= -2\ka r \,\dt u^2 + 2\, \dt u\, \dt r + \varphi(u,r)^\frac{2}{d}\,\dt\O_d^2 , \\
	\varphi(u,r) &\equiv \Phi(u,r)^d, \qquad \Phi(u,r) \equiv  L - \frac{1}{2\ka} e^{\ka u + \a} - r e^{-\ka u - \a} ,
\end{split}
\end{align}
where $\kappa$ is a spacetime constant known as the inaffinity (see Footnote~\ref{fn:inaffinity}), and $\varphi$ parametrizes the area of the transverse sphere at a given point in spacetime and can be viewed as a dilaton. In these new coordinates, constant $r$ hypersurfaces consist of worldlines with constant acceleration, and $u$ is the clock along these worldlines. For the specific case where $r = \frac{1}{2\ka}$, which is the hypersurface comprising worldlines with proper acceleration $a = \kappa$ and Unruh temperature $\beta^{-1}  = \frac{\kappa}{2\pi}$, $u$ corresponds precisely to the proper time. Furthermore, the bifurcate horizon $\CB$, which is defined here to be the intersection of the past and future horizons of the causal diamond, is reached by first taking $r \to 0$ and then $u \to -\infty$. Taking this particular limit, it is clear from \eqref{eqn:metric} that
\begin{align}\label{eq:B}
	\lim_{u \to -\infty} \varphi(u,0) = L^d \,,
\end{align}
which is precisely the size of the causal diamond. We have drawn a spacetime diagram of the causal diamond in Figure~\ref{fig:SH}.

However, note that strictly speaking, the bifurcate horizon $\CB$ is not covered by the Gaussian null coordinates since we need to take $u \to -\infty$. Indeed, there are many ways to reach $\CB$ from the interior of the causal diamond. For instance, consider the hypersurface $\CH_s$ by fixing $r = r_0 > 0$, which as we mentioned above consists of constantly acclerating observer worldlines.\footnote{Such hypersurfaces are also known as stretched horizons, and have been studied extensively in the fluid gravity literature \cite{Damour:1979wya, Price:1986yy, Thorne:1986iy, Susskind:1993if, Parikh:1997ma}. More recently, they also play an instrumental role in the semiclassical analysis of area fluctuations along a stretched horizon in \cite{Ciambelli:2025flo}.} Let $u = u_\pm$ correspond to the top and bottom tips of $\CH_s$, which is defined by the transverse sphere vanishing, {\em i.e.} $\varphi(u_\pm, r_0) = 0$. Using \eqref{eqn:metric}, it is clear this corresponds to
\begin{align}\label{eqn:upm}
	u_\pm = - \frac{\a}{\kappa} + \frac{1}{\ka} \log\bigg[ \ka L \bigg( 1 \pm \sqrt{1 - \frac{2 r_0}{\ka L^2}} \bigg) \bigg] .
\end{align}
Note that while $u_+$ remains finite as $r_0 \to 0$, we have $u_- \to -\infty$ as $r_0 \to 0$. 

Given \eqref{eqn:upm}, we can determine $s_\CC(r_0)$, which we define to be the distance between the tips of the stretched horizon and the top and bottom tips of the causal diamond, where the caustics form (see Figure~\ref{fig:SH}). Using the fact $\tr=0$ at the tips, we obtain the tips of the stretched horizon are located at Minkowski time 
\begin{align}\label{t-dif}
\begin{split}
	\tt_\pm &= - L + \frac{1}{\ka} e^{\ka u_\pm + \a} = \pm L\bigg( 1 - \frac{ r_0}{\ka L^2} \bigg) + \CO(r_0^2),
\end{split}
\end{align}
where we used \eqref{eqn:upm} and Taylor expanded around small $r_0 \ll L$. As the top and bottom tips of the causal diamond are at $\tt = \pm L$, this implies the separation between the top and bottom tips of the causal diamond and the stretched horizon is to leading order in $r_0$ 
\begin{align}\label{Ds-C}
\begin{split}
	s_\CC(r_0) &= \frac{r_0}{\ka L}.
\end{split}
\end{align}

 When we later regularize the on-shell action, it is also useful to determine $u_\mid$, which is the $u$ coordinate of the point on $\CH_s$ that is closest to the bifurcate horizon (see Figure~\ref{fig:SH}). It is straightforward to compute, for the hypersurface $r = r_0$, that the midpoint is given by the coordinate 
\begin{align}\label{eqn:umax}
	u_{\rm mid} = -\frac{\a}{\ka} + \frac{1}{2\ka}\log(2 \ka r_0).
\end{align}
At $u=u_\mid$,  we have from \eqref{eqn:metric}
\begin{align}\label{phi-mid}
	\varphi(u_\mid,r_0) = \bigg( L - \sqrt{\frac{2r_0}{\ka}} \bigg)^d,
\end{align}
which indeed becomes $L^d$ as $r_0 \to 0$. Knowing the value of $u_\mid$, we can also determine $s_\CB(r_0)$, which denotes the separation between the midpoint of the stretched horizon and the bifurcate horizon $\CB$. Substituting \eqref{eqn:umax} into the second equation of \eqref{diffeo-def}, we get
\begin{align}\label{Ds-B}
\begin{split}
	\tr = L - \sqrt{\frac{2r_0}{\ka}}  \quad\implies\quad s_\CB(r_0) = \sqrt{\frac{2r_0}{\ka}} .
\end{split}
\end{align}
We observe this is consistent with \eqref{phi-mid}, which is precisely the area of a sphere with radius smaller by $s_\CB(r_0)$.

Although Gaussian null coordinates are convenient for our computation of the on-shell action, spherical Rindler and topological black hole coordinates are better suited for studying how the geometry responds in the presence of modular fluctuations e.g., see \cite{Verlinde:2019xfb,Verlinde:2019ade}.\footnote{We collected the relationship the various coordinate systems have with each other, as well as the conformal Killing frame, which is useful to physically understand modular flow inside the causal diamond, in Appendix~\ref{app:coordinates}.} Defining
\begin{align}\label{map0}
\begin{split}
	u = T + \frac{1}{2\ka} \log(2\ka r) - \frac{\a}{\ka} , \qquad r = L(1 + 2\z) -R,
\end{split}
\end{align}
where $\z$ is a spacetime constant (but not necessarily a phase space constant, i.e., $\d\z \not= 0$),\footnote{The fact that we do not require $\d\z = 0$ is explained in Appendix~\ref{app:tbh-reg}, below \eqref{Th-int2}.} introduced to match the conventions of \cite{Verlinde:2019xfb, Verlinde:2019ade} and will later be identified with an effective Newtonian potential. Substituting this into \eqref{eqn:metric}, we obtain the topological black hole metric 
\begin{align}\label{eqn:topBH}
\begin{split}
	\dt s^2 &= -2 \ka  f(R)\, \dt T^2  + \frac{\dt R^2}{2 \ka  f(R)}  + \varphi(T,R)^{\frac{2}{d}}\, \dt\O_d^2 \\
	\varphi(T,R) &\equiv \Phi(T,R)^d, \qquad  \Phi(T,R) \equiv  L - \sqrt{\frac{2  f(R)}{\ka}} \cosh\left(\ka T\right) , \qquad f(R) \equiv L(1 + 2\z) - R ,
\end{split}
\end{align}
where we have suggestively defined $f(R)$ to resemble an emblackening factor. By making the further identification 
\begin{align}\label{map}
	r = L(1 + 2\z) - R  =  f(R) = \frac{\ka}{2}\eta^2 ,
\end{align}
and substituting this into \eqref{eqn:topBH}, we obtain the spherical Rindler metric
\begin{align}\label{eq:rindler}
\begin{split}
	\dt s^2 &= - (\ka \eta)^2\, \dt T^2  + \dt \eta^2  + \big( L - \eta \cosh(\ka T) \big)^2\, \dt\O_d^2 \, .
\end{split}
\end{align}
These coordinates will be particularly conducive for using the replica method to construct the $n$-fold cover of the causal diamond, so that we can compute the $n$-R\'enyi entropy for the reduced density matrix associated to the causal diamond. 

Indeed, to use the replica method, we first Wick rotate to Euclidean time by taking $T = -i T_E$, so that the Euclidean version of \eqref{eq:rindler} is
\begin{align}\label{eq:euclidean}
\begin{split}
	\dt s^2 = (\ka \eta)^2 \, \dt T_E^2 + \dt \eta^2 + \big( L - \eta \cos(\ka T_E) \big)^2 \,\dt \O_d^2.
\end{split}
\end{align}
Notice that this metric still has (Euclidean) time dependence in its transverse components. To eliminate this time dependence so that we can appropriately construct the replica manifold, we would like to restrict ourselves to near the bifurcate horizon $\CB$. This corresponds to taking $\eta \to 0$, since in the original Gaussian null coordinate this is equivalent to taking $r \to 0$ and $u \to -\infty$ simultaneously. In this limit, we get to leading order
\begin{align}\label{eq:euclidean-metric}
\begin{split}
	 \dt s^2\big|_{\text{near $\CB$}} = (\ka\eta)^2\, \dt T_E^2 + \dt \eta^2 + L^2 \,\dt\O_d^2.
\end{split}
\end{align}
Thus, we see that the line element decomposes into a direct product betweeen a Euclidean Rindler spacetime in the $(T_E,\eta)$ direction and a transverse $d$-sphere with fixed radius $L$. To ensure this line element describes a smooth geometry, we demand $T_E$ has the periodic identification
\begin{align}\label{eqn:TEperiod}
	T_E \sim T_E + \frac{2\pi}{\ka} ,
\end{align}
so that the Euclidean Rindler spacetime is simply the flat metric in polar coordinates. Note that this is the same Euclidean periodicity which is needed to (minimally) ensure the single-valuedness of the transverse space in the full geometry \eqref{eq:rindler}.

\section{On-shell Action} \label{sec:onshell}

In this section, we will compute an on-shell action associated to the causal diamond. We will then in the next section extend our analysis and compute the analogous on-shell action associated to the replica manifold. This will be instrumental for studying the entanglement properties of the causal diamond, and the fluctuations in the geometry that appear in tandem with the fluctuations in the entanglement. Because we are restricting ourselves to metrics of the form \eqref{eqn:metric} (or equivalently \eqref{eqn:topBH}), we are also restricting the class of perturbations we are considering. Allowing for more general metric fluctuations, such as relaxing spherical symmetry, would be very interesting, and we hope to consider such fluctuations in future work.

To determine the on-shell action, we begin by evaluating the pre-symplectic potential $\widetilde\Th$ of the bulk Einstein-Hilbert action for the family of metrics expressible in Gaussian null form. This is precisely the variation of the bulk action. By imposing appropriate boundary conditions so that $\widetilde\Th$ is a total variation, this allows us to determine the appropriate boundary action $I_\OS$  we can add to the bulk action to ensure the total action is invariant under variations respecting the boundary conditions. In other words, our boundary action $I_\OS$ is determined so that
\begin{align}
	\d I_\OS = - \widetilde \Th.
\end{align}
Since the Einstein-Hilbert action vanishes on-shell, $I_\OS$ is precisely the on-shell action as well.\footnote{The on-shell action for a certain family of shockwave metrics and metrics exhibiting the leading memory effect was determined in this manner recently in \cite{He:2024vlp}.} 

To evaluate the pre-symplectic potential associated to our Minkowski causal diamond, we recall that for a general metric, the pre-symplectic potential on the boundary hypersurface $\Sigma$ is given by \cite{Carroll:2004st}
\begin{align}\label{theta-def}
\begin{split}
	\widetilde\Th_\Sigma = \frac{1}{16\pi G_N} \int_{\Sigma} \dt\Sigma_\mu \,\big( g^{\nu\rho} \delta\G^\mu_{\nu\rho} - g^{\mu\nu}\delta \G^\rho_{\nu\rho} \big) ,
\end{split}
\end{align}
where $\dt\Sigma_\mu$ is the surface element on the boundary $\Sigma$. It is convenient for us to evaluate $\widetilde\Th$ in Gaussian null coordinates, and we are interested in taking $\Sigma$ to be the hypersurface specified by $r=r_0$. For $r_0 >0$, this corresponds to a stretched horizon $\CH_s$, while in the case $r_0 = 0$, the stretched horizon $\CH_s$ becomes the future horizon $\CH^+$ of the causal diamond.\footnote{If we chose ingoing rather than outgoing null coordinates at the beginning, the limit $r \to 0$ would correspond to the past horizon $\CH^-$ instead.} In either case, the measure on the hypersurface is given by\footnote{The negative sign in the measure is due to the fact we oriented our normal vector to be outward-pointing, which is in the direction of \emph{decreasing} $r$ along constant $u$ rays.}
\begin{align}
	\dt\Sigma_\mu = -\delta^r_\mu \varphi(u,r)\,\dt u\,\dt\O_d.
\end{align}
Substituting this into \eqref{theta-def}, we get
\begin{align}\label{theta-def2-app}
\begin{split}
	\widetilde\Th_{\CH_s} = - \frac{1}{16\pi G_N} \int_{\CH_s} \dt u\,\dt\O_d\, \varphi\big( g^{\nu\rho} \d\G^r_{\nu\rho} - g^{r\nu}\d\G^\rho_{\nu\rho} \big) .
\end{split}
\end{align}
Straightforwardly evaluating the relevant Christoffel symbols for the metric \eqref{eqn:metric} and taking their variation, we obtain
\begin{align}\label{eqn:thetastart2rneq0}
\begin{split}
	\widetilde\Th_{\CH_s} &= \frac{\O_d}{8\pi G_N} \int_{\CH_s} \dt u \bigg[ \Phi^d \bigg( \delta \ka + d \frac{\p_u\delta\Phi}{\Phi} \bigg) + r_0 d \Phi^{d-1}\big( \d \ka \p_r\Phi + 2\ka \p_r\d\Phi \big)  \bigg]\bigg|_{r=r_0} ,
\end{split}
\end{align}
where we have trivially carried out the angular integral to pick up the solid angle $\O_d$. Now, the boundary of our causal diamond corresponds to fixing $r_0 = 0$, in which case our boundary corresponds to the future causal horizon $\CH^+$. In this case, the second term in \eqref{eqn:thetastart2} vanishes, and our pre-symplectic potential becomes
\begin{align}\label{eqn:thetastart2}
\begin{split}
	\widetilde\Th^+ &= \frac{\O_d}{8\pi G_N} \int_{u_\CB}^{u_+} \dt u \, \Phi^d \bigg( \delta \ka + d \frac{\p_u\delta\Phi}{\Phi} \bigg)\bigg|_{r=0} ,
\end{split}
\end{align}
where we have demarcated the endpoints of the integral to be $u_\CB$, which is the location of the bifurcate horizon $\CB$, and $u_+$, given explicitly by \eqref{eqn:upm} with $r_0 = 0$ and corresponds to the top tip of the causal diamond. Notice that as was explained around \eqref{eq:B}, formally $u_\CB \to -\infty$, although we will eventually explain how to regularize this divergence.

To further simplify \eqref{eqn:thetastart2}, let us define 
\begin{equation}\label{eqn:mu}
	\mu(u,r) \equiv \frac{1}{2} \log \frac{2 \partial_u \Phi(u,r)}{\partial_r \Phi(u,r)}  = \frac{1}{2} \log \big( e^{2(\ka u + \a )} - 2 \ka r \big) ,
\end{equation}
where in the second equality we used the definition of $\Phi$ given in \eqref{eqn:metric}.\footnote{This is precisely $\mu$ defined in \cite{Bub:2024nan}, except we have extended the definition to the entire causal diamond instead of restricting it to the bifurcate horizon $\CB$.} From \eqref{eqn:mu}, it immediately follows
\begin{align}\label{mu-prop}
\begin{split}
	\p_u\mu(u,0) &= \ka , \qquad \d\mu(u,0) = \d(\ka u + \a ) = \frac{\p_u \d\Phi(u,0)}{\p_u\Phi(u,0)}.
\end{split}
\end{align}
Substituting these relations into \eqref{eqn:thetastart2}, we obtain
\begin{align}\label{eqn:thetastart3}
\begin{split}
	\widetilde{\Theta}^+  &= \frac{\Omega_d}{8 \pi G_N} \int_{u_{\CB}}^{u_+} \dt u \,  \big( \Phi^d \partial_u \delta \mu +  d \Phi^{d-1} \p_u\Phi \delta \mu\big)\Big|_{r = 0}  \\
	&= \frac{\O_d}{8\pi G_N} \int_{u_\CB}^{u_+} \dt u  \,\p_u\big( \Phi^d \delta \mu  \big)\Big|_{r=0} \\
	&= - \frac{\O_d}{8\pi G_N} L^d \d\mu(u_\CB,0) \\
	&= - \frac{A \d\mu_\CB}{8\pi G_N} ,
\end{split}
\end{align}
where in the penultimate equality we used $\Phi^d(u_\CB,0) = L^d$ by \eqref{eq:B} (recall $u_\CB \to -\infty$), and that $\Phi(u_+,0) = 0$ as the transverse metric vanishes at the top and bottom tips of the diamond. In the final equality, we denoted $A \equiv \O_d L^d$ to be the area of the causal diamond, and $\mu_B \equiv \mu(u_\CB,0)$.

To obtain the boundary action to add to the Einstein-Hilbert action so that the total variation vanishes, we need to choose appropriate boundary conditions such that the pre-symplectic potential is a total variation. This is equivalent to choosing either Dirichlet or Neumann boundary conditions in order to obtain a well-defined variational problem. For us, we are interested in first fixing the unperturbed area of the causal diamond.\footnote{Later in Section~\ref{subsec:LT}, we will use the Legendre transform to change the boundary conditions and study what happens when the area is allowed to fluctuate.} This is achieved by fixing
\begin{align}\label{fixed-area}
\begin{split}
	A = A_\CB = \O_d L^d_\CB
\end{split}
\end{align}
for a fixed length $L_\CB$ and area $A_\CB$, so that
\begin{align}\label{}
	\d A = \d A_\CB = 0.
\end{align}
which case we can rewrite \eqref{eqn:thetastart3} as
\begin{align}
\begin{split}
	\widetilde\Th^+ &= -\d \bigg( \frac{A_\CB\mu_\CB}{8\pi G_N} \bigg) .
\end{split}
\end{align}
As this is the variation of the Einstein-Hilbert action, we see that we can cancel this total variation if we add the boundary action
\begin{align}\label{eqn:Son-shell}
	I_\OS = \frac{A_\CB \mu_\CB}{8\pi G_N} .
\end{align}
More explicitly, using \eqref{eqn:mu} with $(u,r) = (u_\CB,0)$, we see that
\begin{align}\label{eqn:Son-shellfixed}
\begin{split}
	I_\OS = \frac{A_\CB}{8\pi G_N} (\ka u_\CB + \a).
\end{split}
\end{align}
This is the unregularized Lorentzian on-shell action associated to the causal diamond, and formally diverges since $u_\CB \to -\infty$. A more careful treatment of regularizing the on-shell action is given in Appendix~\ref{app:gn-reg} and explicitly clarifies the nature of the divergence $u_\CB$.

\section{From Euclidean Path Integral to Thermodynamics}\label{sec:thermo}

As we discussed in the introduction, the density matrix associated to the Rindler wedge of the Minkowski half-space is thermal (with respect to boost time), and hence it has thermal fluctuations. Therefore, near the horizon of a sufficiently large causal diamond, where we can utilize the planar approximation, we also expect the density matrix to be approximately thermal. In particular, this approximation holds near the bifurcate horizon, where the on-shell action localizes. Consequently, by taking derivatives with respect to the inverse temperature $\beta$ of the partition function, as in \eqref{eq:thermK}, we obtain $\langle K \rangle = S$ and its fluctuations $\la (\Delta K)^2\ra$. Thus, we would like to determine what is $\log \CZ$ as a function of $\b$. As emphasized in \eqref{eqn:TEperiod}, smoothness of the metric demands that the Euclidean time be periodic with period $\frac{2\pi}{\ka}$.  As the inverse temperature is identified with the Euclidean time periodicity, this implies that the inverse temperature of the causal diamond (near the bifurcate horizon) is precisely $\beta = \frac{2\pi}{\kappa}$. Because the replica index changes the Euclidean time periodicity, so that
\begin{equation}
	\beta \to \beta n,
\label{eq:replica}
\end{equation}
we can view variations in $\b$ equivalently as variations in $n$.

We can now use the replica trick to compute the mean and variance of the modular Hamiltonian. First, we recall \eqref{Z-path3-intro} from the introduction, which for convenience we reproduce here:
\begin{align}\label{Z-path3}
\begin{split}
	\CZ[n] \equiv \Tr\,\rho^n , \qquad \log\CZ[n] = - I^\reg_\OS[n].
\end{split}
\end{align}
This equation derives from the fact the gravitational path integral over the $n$-fold causal diamond, which can be semiclassically approximated by the exponential of our $n$-fold replica on-shell action, computes $\Tr\,\rho^n$.  We will explain how to obtain and regularize the replica on-shell action in Section~\ref{subsec:replica}, but we first turn to how to use $\CZ[n]$ as a generating function to compute $\la K \ra$ and $\la (\D K)^2 \ra$.

When utilizing the replica trick, we are effectively in a canonical ensemble, with replica index $n$ playing the role of the inverse temperature $\b$ and $K$ playing the role of the Hamiltonian.\footnote{In a canonical ensemble, the partition function is $\CZ_\b = \Tr\, e^{-\b H} = \sum_E e^{-\b E}$. We can then compute the mean and variance of the energy at any fixed inverse temperature $\b$ by taking appropriate derivatives of $\CZ_\b$ with respect to $\b$, namely $\la E \ra = -\p_\b \log \CZ_\b$ and $\la (\D E)^2 \ra = \p^2_\b \log \CZ_\b$. \label{fn:canonical}} We briefly review the replica trick here, though it is described extensively in the literature, e.g. see \cite{Lewkowycz:2013nqa, Dong:2013qoa, Dong:2016fnf} and references therein, and we refer the reader to those papers for more details. First, note that the $n$-R\'enyi entropy $S[n]$ is
\begin{align}\label{Sn}
\begin{split}
	S[n] &\equiv \frac{1}{1-n}\log\Tr\,\hat \rho^n = \frac{1}{1-n} \big(\log\CZ[n] - n\log\CZ[1] \big) , \qquad \hat\rho \equiv \frac{\rho}{\CZ[1]}.
\end{split}
\end{align}
The replica trick states that the von Neumann entropy is given by $S = \lim_{n \to 1} S[n]$, in which case using \eqref{Sn} we obtain
\begin{align}
\begin{split}
	S = \lim_{n \to 1} \frac{1}{1-n} \big( \log\CZ[n] - n \log \CZ[1] \big) = -\lim_{n \to 1} (\p_n - 1) \log \CZ[n] .
\end{split}
\end{align}
Recalling that $K \equiv -\log\hat\rho$, we obtain
\begin{align}\label{K-1pt}
\begin{split}
	\la K \ra_{n=1} = S = -\lim_{n \to 1} (\p_n - 1) \log \CZ[n] ,
\end{split}
\end{align}
where we use the subscript $n=1$ to emphasize the fact the expectation value is taken for fixed $n=1$. 

Next, we compute the variance of $K$, which is much less explored in the literature. We begin by computing
\begin{align}\label{K-2-int}
\begin{split}
	\la K^2 \ra_{n=1} = \lim_{n \to 1} \Tr\big(e^{-n K} K^2 \big) = \lim_{n \to 1} \p_n^2 \Tr(e^{-nK}) = \lim_{n \to 1} \p_n^2 \bigg( \frac{\CZ[n]}{\CZ[1]^n} \bigg) .
\end{split}
\end{align}
This expression is rather compact, but it will be more useful for us to further evaluate it to obtain
\begin{align}\label{K-2-int2}
\begin{split}
	\la K^2 \ra_{n=1} &= \lim_{n \to 1} \p_n^2 \Big( \CZ[n] e^{-n\log\CZ[1]} \Big) \\
	&= \lim_{n \to 1} \frac{1}{\CZ[1]} \Big( \p_n^2 \CZ[n] - 2(\log\CZ[1])\p_n\CZ[n] + \CZ[1](\log\CZ[1])^2 \Big).
\end{split}
\end{align}
From \eqref{K-1pt} and \eqref{K-2-int2}, it follows the variance of $K$ is given by
\begin{align}\label{DK2}
\begin{split}
	\la (\D K)^2 \ra_{n=1} &\equiv \la K^2 \ra_{n=1} - \la K \ra^2_{n=1} = \lim_{n \to 1} \p_n^2\log \CZ[n].
\end{split}
\end{align}

Our goal is now to compute $I_\OS[n]$ for the replica manifold and regularize it, so that we may use it to compute the mean and variance of $K$. In Section~\ref{subsec:replica}, we will construct the replica manifold and its associated regularized Euclidean on-shell action. In Section~\ref{subsec:EE}, we will compute the mean and variance of $K$. Finally, we explore the statistics associated to fluctuations of the replica index for fixed $K$ in Section~\ref{subsec:LT} via a Legendre transform.

\subsection{On-shell Action for the Replica Manifold}\label{subsec:replica}

The replica manifold is most straightforwardly constructed using spherical Rindler coordinates near the bifurcate horizon, whose metric to leading order is given by \eqref{eq:euclidean-metric}. The discussion around \eqref{eq:replica} makes it clear that a replica variation corresponds to changing the Euclidean time periodicity from $\frac{2\pi}{\ka}$ to $\frac{2\pi n}{\ka}$, so that the new line element is
\begin{align}\label{eq:rindler-NH-n}
	\dt s^2_{(n)}\big|_{\text{near horizon}} &= \frac{(\ka \eta)^2}{n^2} \, \dt T_E^2 + \dt \eta^2 + L^2 \, \dt \O_d^2.
\end{align}
To extend this metric away from the near horizon limit, we first analytically continue $T_E = iT$ back to Lorentzian time and then impose Einstein's equations. This is simply \eqref{eq:rindler} with $\ka \to \frac{\ka}{n}$, so that the line element associated to the replica manifold becomes\footnote{Recall from the discussion below \eqref{eq:euclidean} that we approach the bifurcate horizon $\CB$ by taking $\eta \to 0$.}
\begin{align}\label{eq:rindler-n}
\begin{split}
	\dt s^2_{(n)} &= - \frac{(\ka \eta)^2}{n^2} \, \dt T^2  + \dt \eta^2  + \bigg( L - \eta \cosh \bigg( \frac{\ka T}{n} \bigg) \bigg)^2 \,\dt\O_d^2 .
\end{split}
\end{align}
Interestingly, since $\kappa$ enters into the metric, fluctuations in $n$, which correspond to thermodynamic variations, translate into geometric variations.

We can rewrite the replica manifold metric in Gaussian null coordinates by inverting the coordinate transformations \eqref{map0} and \eqref{map}, with the replacement $\ka \to \frac{\ka}{n}$. The inverted coordinate transformations are thus given by
\begin{align}\label{eqn:RindlertoGNn}
\begin{split}
	T &= u - \frac{n}{2\ka} \log\bigg( \frac{2\ka r}{n} \bigg) + \frac{\a n}{\ka} , \qquad \eta = \sqrt{\frac{2nr}{\ka}},
\end{split}
\end{align}
and substituting this into \eqref{eq:rindler-n}, we obtain
\begin{align}\label{gaussian-met-n}
\begin{split}
	\dt s^2_{(n)} &= - \frac{2\ka r}{n}\, \dt u^2 + 2\, \dt u\,\dt r + \varphi_n(u,r)^{\frac{2}{d}}\, \dt \O_d^2 \\
	\varphi_n(u,r) &= \bigg(L - \frac{n}{2\ka} e^{\frac{\ka u}{n} + \a} - re^{-\frac{\ka u}{n} - \a} \bigg)^d .
\end{split}
\end{align}
In particular, we see that near the bifurcate horizon, the apparent effect of the $n$-fold copy of the Gaussian null metric \eqref{eqn:metric} associated to a causal diamond is to simply take $\ka \to \frac{\ka}{n}$ in our original line element. Thus, the on-shell action in Gaussian null coordinates associated to the replica manifold \eqref{gaussian-met-n} immediately, and trivially, leads to the on-shell action
\begin{equation}\label{eqn:Son-shellfixedn}
	I_\OS[n]=  \frac{A_{\CB}}{8 \pi G_N}\left(\frac{\kappa}{n} u_\CB + \alpha\right) .
\end{equation}
We remark that the on-shell action's dependence on $n$, or equivalently the inverse temperature, was also obtained for the gravitational effective actions given in \cite{Nakaguchi:2016zqi, Harlow:2018tqv}. 

In order to regularize \eqref{eqn:Son-shellfixedn}, we begin by considering what is the finite on-shell action computed on a stretched horizon at $r = r_0 > 0$. The computation is more involved and is relagated to Appendix~\ref{app:gn-reg}, and the result is given in \eqref{app:reg-gn-os2} to be
\begin{equation}\label{eqn:Son-shellfixednreg1}
I_\OS[n]=  \frac{A_{\CB}}{8 \pi G_N}\left(\frac{\kappa}{n} u_{\text{reg}} + \alpha\right)  ,
\end{equation}
where $u_\reg$ is a point on the stretched horizon that we are choosing to be the endpoint limiting to $u_\CB$ when $r_0 \to 0$. As $\CB$ is located at the midpoint between the top and bottom tips of the causal diamond, it is natural to choose $u_\reg = u_\mid$, which is given in \eqref{eqn:umax}. This implies
\begin{equation}\label{eqn:Son-shellfixednreg2}
	I_\OS[n,r_0]=  \frac{A_{\CB}}{8 \pi G_N}\bigg[ \frac{1}{2 n}\log(2 \ka r_0) + \left(1-\frac{1}{n}\right)\alpha\bigg]  .
\end{equation}
We next observe that thus far, we have done a Lorentzian computation. However, it is known from a variety of contexts and examples that the Euclidean thermodynamic partition function can be extracted from such a Lorentzian calculation by considering an appropriate analytic continuation or, equivalently, a complex contour \cite{Colin-Ellerin:2020mva,Colin-Ellerin:2021jev, Banihashemi:2024weu}. The imaginary pieces that are picked up along branch cuts are then precisely the contributions that are not merely pure phases and can reproduce the Euclidean results. Inspired by the prescription advocated in \cite{Glorioso:2018mmw}, which in practice amounts to analytically continuing $r_0 \to r_0 e^{-2 \pi i}$ in an asymptotically AdS spacetime,\footnote{We analytically continue $r_0$ using $e^{-2\pi i}$ rather than $e^{2\pi i}$ because $r_0$ decreases as we move outwards.} we perform the same analytic continuation in \eqref{eqn:Son-shellfixednreg2} and keep only the imaginary piece  contributing to the monodromy of the logarithm. Doing so, we find that the regularized \emph{Euclidean} on-shell action, which is $-i$ times the Lorentzian action, has real contribution given by
\begin{align}\label{onshell-reg-n}
\begin{split}
	I_\OS^\reg[n]  =  - \frac{A_\CB}{8G_N n} .
\end{split}
\end{align}
We did an analogous computation of the regularized Euclidean on-shell action using topological black hole coordinates in Appendix~\ref{app:tbh-reg} and obtained the same result. Interestingly, in that computation, we get a purely imaginary Lorentzian action after performing the analytic continuation, resulting in a real Euclidean action upon Wick rotating.
It would be interesting to better understand the implications of the above prescription given in \cite{Glorioso:2018mmw} for Minkowski causal diamonds, in particular in the context of the Schwinger-Keldysh formalism \cite{Schwinger:1960qe, Keldysh:1964ud} that motivated it. We suspect this is connected to the question of how to construct the density matrix of the causal diamond. Furthermore, it would prove more satisfactory to also obtain a finite Lorentzian action \eqref{eqn:Son-shellfixedn} to begin with by systematically adding boundary counterterms. For our present purposes, however, we restrict ourselves to exploring the implications of the regularized action \eqref{onshell-reg-n}, and leave a more complete treatment of the on-shell action for future work.

\subsection{Entanglement Entropy and Modular Fluctuations}\label{subsec:EE}

Given the regularized Euclidean on-shell action \eqref{onshell-reg-n} for the replica manifold, we can now use the identification $\log \CZ[n] = -I_\OS^\reg[n]$ given in \eqref{Z-path3} to compute the mean and variance of $K$. First, using \eqref{K-1pt}, we compute the mean to be
\begin{align}\label{eqn:Sform}
\begin{split}
	K_\CB \equiv \la K \ra_{n=1} = - \lim_{n \to 1} (\p_n - 1) \bigg( \frac{ A_{\CB}}{8 G_N n} \bigg) = \frac{ A_{\CB}}{4 G_N} = -2I_\OS^\reg[1] .
\end{split}
\end{align}
This is the statement that semiclassically, the entanglement entropy across a Minkowski causal diamond with no matter saturates the covariant entropy bound \cite{Bousso:1999xy}, a result that has been previously obtained using other means for spherically symmetric finite causal diamonds, e.g., see \cite{Jacobson:2015hqa, Balasubramanian:2013rqa, Banks:2020tox}. 

Similarly, we next use \eqref{DK2} to compute the variance of $K$ to be
\begin{align}\label{eqn:Sform2}
\begin{split}
	\la (\D K)^2 \ra_{n=1} &= -\lim_{n \to 1} \p_n^2 \bigg( \frac{\l A_\CB}{8 G_N n} \bigg) = \frac{A_{\CB}}{4 G_N} .
\end{split}
\end{align}
We thus obtain the equality
\begin{align}\label{eqn:VZrelation}
	\la K \ra_{n=1} = \la (\D K)^2 \ra_{n=1} \, ,
\end{align}
a result that was previously obtained in flat spacetimes using other methods in \cite{Nakaguchi:2016zqi,Banks:2021jwj, Gukov:2022oed,He:2024vlp}.\footnote{This relation was also obtained for boundary-anchored causal diamonds in AdS/CFT in \cite{DeBoer:2018kvc,Verlinde:2019ade}.}  
The importance of the relation between the expectation value of the modular Hamiltonian and its fluctuations \eqref{eqn:VZrelation} has been emphasized in \cite{Verlinde:2019xfb,Verlinde:2019ade,Banks:2021jwj, He:2024vlp}. We will discuss in Section~\ref{geometry} geometric implications of these modular fluctuations.

We conclude this subsection with the following observation. Thus far, we have introduced $n$ to be an integer that parametrizes the number of copies of the original manifold we are gluing together, thereby obtaining the metric \eqref{eq:rindler-NH-n}. This is indeed the standard interpretation of the replica index, whereby when performing the replica trick, we would have to analytically continue $n$ to non-integer values. However, it should be clear in our setup that there are no obstructions to $n$ being non-integer. In Gaussian null coordinates, the line element of the replica manifold is given by \eqref{gaussian-met-n}, so that the induced metric on any stretched horizon $r= r_0 >0$ is
\begin{align}
\begin{split}
	\dt s^2_{(n)}\Big|_{\CH_s} &= -\frac{2\ka r_0}{n} \,\dt u^2 + \varphi_n(u,r_0)^\frac{2}{d} \, \dt\O_d^2 .
\end{split}
\end{align}
As was observed in Section~\ref{subsec:replica}, near the bifurcate horizon, the only change to the line element where $n=1$ is that $\ka$ is shifted to $\frac{\ka}{n}$. Recalling \eqref{Ds-B}, we see that the separation between the stretched horizon and the bifurcate horizon becomes under the replica variation
\begin{align}
\begin{split}
s_\CB &= \sqrt{ \frac{2r_0}{\ka} } \to \sqrt{\frac{2 n r_0}{\ka} }.
\end{split}
\end{align}
In other words, for a fixed $r_0 > 0$, we see that after introducing the replica index, the stretched horizon has shifted, with the size of the shift being given by
\begin{align}\label{Ds-nshift}
\begin{split}
	\D s_\CB &\equiv \sqrt{\frac{2n r_0}{\ka}} - \sqrt{\frac{2r_0}{\ka}} =  \frac{\D n}{2}\sqrt{\frac{2r_0}{\ka} } + \CO((\D n)^2) ,
\end{split}
\end{align}
where we used the definition $\D n = n-1$. Thus, we see that instead of having the replica index $n$ parametrize the $n$-fold copy of the causal diamond, we can equivalently interpret it as reparametrizing the inaffinity $\ka \to \frac{\ka}{n}$, so that the stretched horizon at some fixed $r_0$ gets shifted for small $\D n$ by an amount given by \eqref{Ds-nshift}. From this viewpoint, taking $\D n$ small simply corresponds to exploring how $\CH_s$ changes infinitesimally under fluctuations in $n$. Indeed, our interpretation of $\D n$ corresponding to a fluctuation in the stretched horizon, which we affectionately dub a Rindler wobble, can be thought of as a fluctuation in the temperature experienced by accelerating observers living on $\CH_s$. This resonates well with the fact that our $n$ fluctuations is analogous to temperature fluctuations in a thermal ensemble.

\subsection{Replica Variations via Legendre Transform}\label{subsec:LT}

In the previous subsection, we derived the mean and the variance of the modular Hamiltonian in an ensemble where we fixed $n=1$ and $K$ was treated as a random variable allowed to fluctuate. This is analogous to the canonical ensemble, where we are able to compute the mean and variance of the energy, with the energy being the random variable and the temperature being fixed. In this subsection, we now want to trade the role of $n$ and $K$ by treating $n$ as a random variable and study its fluctuations in an ensemble where we fix $K$ to be its expectation value given in \eqref{eqn:Sform}. Utilizing the thermodynamics analog, this is the same as going to the microcanonical ensemble, where we allow the temperature to fluctuate while the energy is fixed instead. In particular, this allows us to study the implications of fluctuations in the replica index. 

The natural way in thermodynamics to go from the canonical to microcanonical ensemble is via a Legendre transform. However, we will first sketch a quick way to obtain the size of the fluctuations in $n$ in the microcanonical ensemble. Begin by Taylor expanding the on-shell action \eqref{onshell-reg-n} about $n=1$, so that we get at second order
\begin{align}\label{taylor}
\begin{split}
	I^\reg_\OS[n] &= I^\reg_\OS[1] + \D n \p_n I^\reg_\OS[n]\big|_{n=1} + \frac{1}{2}(\D n)^2 \p_n^2 I^\reg_\OS[n]\big|_{n=1} + \CO((\D n)^3) \\
	&= n I^\reg_\OS[1] + K_\CB \D n  - \frac{1}{2}(\D n)^2 \la (\D K)^2 \ra_{n=1} + \CO((\D n)^3) ,
\end{split}
\end{align}
where we have defined $\D n \equiv n-1$, and in the second equality we used \eqref{eqn:Sform} and \eqref{eqn:Sform2}. Introducing (e.g., see \cite{Verlinde:2019ade})
\begin{align}\label{F-def}
	\CF[n]  \equiv I^\reg_\OS[n] - n I^\reg_\OS[1],
\end{align}
we can rewrite \eqref{taylor} as
\begin{align}\label{F-def2}
\begin{split}
\CF[n]  &= K_\CB \D n  - \frac{1}{2}(\D n)^2 \la (\D K)^2 \ra_{n=1} + \CO((\D n)^3) .
\end{split}
\end{align}
It is clear from the above equation that $\CF$ is the generating function for the moments of the modular Hamiltonian, since
\begin{equation}\label{eqn:FtoK}
	\la K \ra_{n=1} = \p_n \CF[n]\big|_{n = 1}  , \qquad 	\la (\Delta K)^2 \ra_{n=1} = - \p_n^2 \CF[n]\big|_{n = 1}  .
\end{equation}
This is analogous to the role played by the product $\beta F_{\beta}$, with $F_\b$ being the free energy defined in \eqref{eqn:Fn}, as $\b F_\b$ is by \eqref{eq:thermK} a generating function for moments of $K$ as well. However, due to the constant offset in computing $\la K \ra_{\b}$ in \eqref{eqn:Fn}, $\b F_\b$ with $\b =1$ coincides with $\CF[n]$ only if $F_{\b=1} = 0$, which is equivalent to demanding $\CZ_{\b=1} = 1$. This implies if we define a ``modular free energy'' analogous to \eqref{eqn:Fn}, namely
\begin{align}\label{eq:Fn2}
	F[n] \equiv - \frac{1}{n} \log \CZ[n],
\end{align}
we would have $n F[n] = \CF[n]$ near $n=1$ only if $\CZ[1] = 1$. In this case, by \eqref{eq:Fn2} we can view $e^{-\CF[n]- \D n K_\CB}$ near $n=1$ as a probability density for the random variable $n$ that has not yet been integrated over in the Euclidean path integral with a source $K_\CB$. We see from \eqref{F-def2} that to quadratic order this distribution is Gaussian, with the first and second moments of $\D n$ given by
\begin{align}\label{target}
\begin{split}
	\la \Delta n \ra_{K=K_{\CB}} = 0 , \qquad \la ( \D n)^2 \ra_{K=K_{\CB}} = \frac{1}{\la (\D K)^2\ra}_{n=1} ,
\end{split}
\end{align}
where the subscript $K_\CB$ on the left-hand side of the above equation reminds us the expectation values are taken at $K=K_\CB$. Note that including the source $ K_{\CB}$ in the probability density is required for consistency, as the moments of $K$ are evaluated at $\D n= 0$, and so we would like our probability distribution to reproduce $\la \D n \ra_{K=K_\CB} = 0$. However, a more systematic method of obtaining \eqref{target}, where we are free to fix a source $K$ while allowing $n$ to fluctuate, is to implement a Legendre transform, which we now turn to.

In \eqref{eqn:FtoK}, $\CF[n]$ is a generating function for computing moments $K$, a random variable, at fixed $n=1$. As we would like to have $n$ be the random variable instead now and get a generating function for computing moments of $n$ at fixed $K = K_\CB$, we need to perform a Legendre transform of $\CF[n]$. To this end, first recall that our regularized Euclidean on-shell action as a function of the replica index $n$ is given in \eqref{onshell-reg-n} to be 
\begin{align}\label{I-OS-reg2}
	I_{\OS}^\reg[n] = - \frac{\l A_\CB}{8G_N n}  = - \frac{K_\CB}{2n},
\end{align}
where in the last equality we used \eqref{eqn:Sform}. Substituting this into \eqref{F-def}, we get
\begin{align}\label{CF-def}
	\CF[n]  = - \frac{K_\CB}{2n} + \frac{K_\CB}{2} n.
\end{align}
From \eqref{F-def2}, we see that the conjugate variables appearing on the right-hand side are $\D n$ and $K$. Therefore, the Legendre transform of the free energy is\footnote{To determine that the Legendre transform \eqref{legendre-1} involves the supremum and not the infimum, we first show that $\CF[n]$ is concave:
\begin{align*}
	\p_n^2 \CF[n] &= -\frac{\la K \ra}{n^3} < 0 .
\end{align*}
This implies that $\CF[n] - (\D n) K$ is also a concave function, meaning that this expression has a supremum rather than an infimum.} 
\begin{align}\label{legendre-1}
\begin{split}
	\widetilde \CF[K] &= \sup_{n} \big( \CF[n] - K\D n \big) \\
	&= \sup_{n} \bigg( -\frac{K_\CB}{2 n} + \frac{K_\CB}{2} n - K \D n \bigg) .
\end{split}
\end{align}
To determine what the value of $n$ should be, we require
\begin{align}\label{n-star}
\begin{split}
	&\p_n \bigg( -\frac{K_\CB}{2 n} + \frac{K_\CB}{2} n - K \D n \bigg)\bigg|_{n = n_\star} = 0 \\
	\implies\quad & n_\star = \sqrt{\frac{K_\CB}{K_\CB + 2 \D K} }, \qquad \D K \equiv K - K_\CB. 
\end{split}
\end{align}
It is easy to check that $n_\star$ is the supremum rather than the infimum. Furthermore, for small $\D K$, we can approximate the above expression with
\begin{align}\label{n-star1}
\begin{split}
	n_\star = 1 - \frac{\D K}{K_\CB} + \frac{3}{2}\frac{(\D K)^2}{K_\CB^2} + \CO((\D K)^3) . 
\end{split}
\end{align}
Substituting \eqref{n-star1} into \eqref{legendre-1} and writing $K = K_\CB + \D K$, we get after some straightforward algebra
\begin{align}\label{CF-legendre-fin}
\begin{split}
	\widetilde\CF[K] &=  K - K_{\CB}\sqrt{1+ \frac{2\D K}{K_{\CB}}} =  \frac{(\D K)^2}{2K_\CB}  + \CO( (\D K)^3) .
\end{split}
\end{align}

We can now use \eqref{CF-legendre-fin} as a generating function to compute the moments of $\D n$. The first moment of $\D n$ is given by 
\begin{align}\label{Dn-1}
\begin{split}
	\la \D n \ra_{K=K_\CB} &= - \p_K \widetilde \CF[K] \bigg|_{K = K_\CB} = - \frac{\D K}{K_\CB}\bigg|_{K = K_\CB} = 0 ,
\end{split}
\end{align}
where the minus sign arises from the sign of $\D n$ in \eqref{legendre-1} when implementing the Legendre transform, and in the last equality we recalled the definition $\D K \equiv K - K_\CB$ from \eqref{n-star}. Furthermore, the second moment is
\begin{align}\label{Dn-2}
\begin{split}
	\la (\D n)^2 \ra_{K=K_\CB} &= \p_K^2  \widetilde\CF[K] \bigg|_{K = K_\CB} = \frac{1}{K_\CB}  = \frac{1}{\la (\D K)^2 \ra_{n=1}} ,
\end{split}
\end{align} 
where we used \eqref{eqn:VZrelation}. Thus, we see that \eqref{Dn-1} and \eqref{Dn-2} are precisely \eqref{target}, as promised.

Finally, as we already mentioned, we can view our above procedure to obtain \eqref{Dn-2} from a standard thermodynamic perspective as going from a canonical ensemble with fixed ``temperature'' $\frac{1}{n}$ to a microcanonical ensemble with fixed ``energy'' $K$. Recalling the thermodynamic identity
\begin{equation}
	S(E) =  \beta E - \beta F_{\beta}  ,
\end{equation}
this means the Legendre transform of the free energy is the microcanonical entropy, which is the logarithm of the density of energy eigenstates. In our case, we can compute
\begin{align}
\begin{split}
	\CS(K) &\equiv n_\star K - \CF[n_\star] =  K_\CB \sqrt{1 + \frac{2 \D K}{K_\CB} } ,
\end{split}
\end{align}
where we substituted in \eqref{CF-def} and \eqref{n-star}. This quantity can be interpreted to count the density of states of fixed modular energy, and the fact that the density involves the square root of $\D K$ is closely related to the relation \eqref{eqn:VZrelation}, as was argued in detail in \cite{Banks:2021jwj}, which equivalently implies $\log \CZ[n] \propto \frac{1}{n}$ (e.g., see \cite{Harlow:2018tqv,Gukov:2022oed}).

\section{Fluctuations in the Geometry}\label{geometry}

We have thus far been focused on the thermodynamic aspects of modular fluctuations in the previous section. We would now like to understand the geometric implications of such fluctuations. In Section~\ref{subsec:fluc-horizon}, we will analyze how the horizon location changes under modular fluctuations and reproduce the results of \cite{Verlinde:2019xfb}. We then compute in Section~\ref{subsec:photon-fluc} a physical observable, namely the phase differences of photons when traversing a stretched horizon versus the causal horizon.

\subsection{Fluctuations of the Stretched Horizon}\label{subsec:fluc-horizon}

To understand how the geometry responds to the fluctuations computed in the previous section, we explore how the causal horizon changes as $n \not=1$ and $K \not= K_\CB$, and what are some possible observational implications. As discussed in the previous section, the choice of ensemble will determine whether it is $n$ or $K$ that is allowed to fluctuate. While we have discussed explicitly how $n$ enters into the geometry, as is evidenced by \eqref{eq:rindler-n}, we have not yet done so for $K$. For this reason, we will denote the metric in this subsection as $g_{\mu\nu}[n,K]$ to keep explicit the dependence the metric has on both $n$ and $K$.

Before making explicit the metric $g_{\mu\nu}[n,K]$ in our setup, it is important to emphasize that when considering spacetime fluctuations, it is not sufficient to simply express $g_{\mu\nu}[n,K]$ as a function of the variation parameters $n$ and $K$ in some coordinate system. This is because implicitly, the coordinate system itself can vary with $n$ and $K$, and so under such fluctuations the metric may correspond to coordinates on different manifolds. It follows that we can compare metrics, and for instance compute 
\begin{align}
	h_{\mu\nu}[n,K] \equiv g_{\mu\nu}[n',K'] - g_{\mu\nu}[n,K] ,
\end{align}
only after a relation is established between the two manifolds on which the metrics are defined. If we change the coordinate identifications between the two manifolds, say by shifting $x^\mu \to x^\mu + \xi^\mu$, this would correspond to at a linearized level the usual linearized gauge transform $h_{\mu\nu} \to h_{\mu\nu} - \nabla_\mu \xi_\nu - \nabla_\nu \xi_\mu$. Naturally, we require physical observables to be independent of such gauge choices.

As an explicit example, consider a causal diamond in Minkowski spacetime described by a different coordinate system from $(\tt,\tr)$, with the line element
\begin{align}\label{new-mink}
	\dt s^2 &= -\dt \tilde\tau^2 + \dt \tilde\rho^2 + \tilde\rho^2\,\dt\O_d^2,
\end{align}
centered at the origin and bounded by causal horizons given by $\tilde\tau = \pm ( L_\CB - s_\CB - \tilde\rho)$, where $s_\CB$ is given by \eqref{Ds-B} with $r_0 = \D L$, so that
\begin{align}\label{Ds-B2}
	s_\CB \equiv s_\CB(\D L) = \sqrt{\frac{2\D L}{\ka}}.
\end{align}
Comparing with Figure~\ref{fig:SH}, it is clear this corresponds to a causal diamond with the bifurcate horizon given at the midpoint of the stretched horizon. If we want to describe the causal boundaries in this different coordinate system in terms of our original $(\tt,\tr)$ coordinates, we need to define how to relate the two metrics. One possibility is the coordinate relations given by
\begin{align}\label{example-map}
\begin{split}
	\tilde\tau &= \begin{cases}
		\sqrt{\tt^2 + s_\CB^2}  - s_\CB & \tilde\tau > 0 \\
		s_\CB - \sqrt{\tt^2 + s_\CB^2} & \tilde\tau \leq 0
	\end{cases} \\
	\tilde\rho &= \tr.
\end{split}
\end{align}
With this identification, note that the future boundary of the causal diamond in the $(\tilde\tau,\tilde\rho)$ coordinate system is mapped to
\begin{align}\label{future-map}
\begin{split}
	\tilde\tau = L_\CB - s_\CB - \tilde\rho \quad &\implies\quad \tt^2  = (L_\CB - \tr)^2 - s_\CB^2,
\end{split}
\end{align}
and similarly the past horizon is mapped to 
\begin{align}\label{past-map}
\begin{split}
	\tilde\tau = \tilde\rho - (L_\CB - s_\CB) \quad &\implies\quad \tt^2  = (L_\CB - \tr)^2 - s_\CB^2.
\end{split}
\end{align}
We now claim that in both cases, the causal horizons are mapped to a stretched horizon in $(\tt,\tr)$ coordinates. To see why, 
we use \eqref{diffeo-def} and the fact the stretched horizon is located at $r = \D L$ to derive that it is parametrized via the equation
\begin{align}\label{eqn:Rindlerrint}
\begin{split}
	\tr = -\frac{\tu}{2} + \frac{L}{2} - \frac{\D L}{\ka ( \tu + L_\CB)} \quad&\implies\quad \tt^2 = (L_\CB - \tr)^2 - s_\CB^2 ,
\end{split}
\end{align}
where in the implication we have used \eqref{Ds-B2}. This is precisely the final equations in \eqref{future-map} and \eqref{past-map}, completing our demonstration that we can map the past and future causal horizons of the smaller causal diamond in $(\tilde\tau,\tilde\rho)$ coordinates to the stretched horizon shown in Figure~\ref{fig:SH} in $(\tt,\tr)$ coordinates.

Importantly, given \eqref{example-map}, the line element \eqref{new-mink} is 
\begin{align}\label{eqn:flatmetric}
\begin{split}
	\dt s^2 &= - \frac{\tt^2}{\tt^2 + s_\CB^2} \dt\tt^2 + \dt\tr^2  + \tr^2\,\dt\O_d^2.
\end{split}
\end{align}
This is certainly not the Minkowski flat metric in $(\tt,\tr)$ coordinates, and we can view it as a perturbed metric with metric perturbation
\begin{align}\label{eqn:hexpl}
\begin{split}
	h_{\tt\tt} = 1 - \frac{\tt^2}{\tt^2 + s_\CB^2} = \frac{s_\CB^2}{\tt^2 + s_\CB^2} ,
\end{split}
\end{align}
and all other components of the metric perturbation vanishing. Thus, this simple example demonstrates how a null horizon in Minkowski spacetime can become a stretched horizon under a metric perturbation.

With this picture in mind, we now proceed to understand the changes in geometry arising from $n$ and $K$ fluctuations. Following \cite{Verlinde:2019xfb,Verlinde:2019ade}, we begin with topological black hole coordinates, where $\ka$ is fixed to $\frac{1}{2L}$ and the emblackening factor $f(R)$ introduced in \eqref{eqn:topBH} is given by
\begin{align}\label{eqn:shiftTPB}
\begin{split}
	f(R) = L[n,K] - R[n,K] + 2\z[n,K] L[n,K] ,	
\end{split}
\end{align}
with $\z$ being identified in \cite{Verlinde:2019xfb, Verlinde:2019ade} as an effective Newtonian potential. In the above equation, we have explicitly indicated that both $L$ and $\z$, as functions on the phase space of causal diamonds, should be expressible in terms of $n$ and $K$. The horizon is located at $f(R_h) = 0$, implying
\begin{align}\label{eqn:Rhorizon}
	R_h[n,K] = L[n,K] + 2 \z[n,K] L[n,K].
\end{align}
Notice that we even wrote the horizon location as $R_h[n,K]$, as it is not a priori true that the radial coordinate is the same for different $n$ and $K$. However, the key identification we will now make, consistent with our derivation of the on-shell action,\footnote{As remarked in Appendix~\ref{app:tbh-reg}, having $\z[n,K]$ be a function in phase space does not affect the on-shell action calculation in topological black hole coordinates.} in order to understand changes in the geometry is to assume
\begin{align}\label{R-relation}
	R[n,K] = R[1,K_\CB].
\end{align}
This allows us to relate any changes to the geometry to the unperturbed reference geometry given by $n=1, K= K_\CB, L[1,K_\CB] = L_\CB$, and $\z[1,K_\CB] = 0$. In particular, it immediately follows from setting $R = R_h$ in \eqref{R-relation} and then using \eqref{eqn:Rhorizon} that\footnote{We remark that we have thus far assumed that $\D L > 0$, as our coordinate system only describes the interior of the causal diamond. However, from symmetry we expect fluctuations in $L$ can be either larger or smaller than $L_\CB$ with equal probability. As we are currently exploring the statistics of modular fluctuations, we will assume henceforth that $\D L$ can be either positive or negative. \label{fn:DL}}
\begin{align}\label{eqn:dLtophi}
	2\z = \frac{L_\CB - L}{L} \equiv \frac{\D L}{L}.
\end{align}
Given the identification
\begin{equation}\label{eqn:Ktemp}
	K = \frac{A}{4G_N} = \frac{\O_d L^d}{4G_N}  ,
\end{equation}
we immediately derive
\begin{align}\label{eqn:dK}
\begin{split}
	\frac{\D K}{K} = \frac{\D A}{A} = \frac{\D L d}{L} + \CO((\D L)^2) = 2\z d + \CO((\D L)^2),
\end{split}
\end{align}
where the final equality follows from \eqref{eqn:dLtophi}. This in turn implies to leading order
\begin{align}\label{eqn:dKtophi}
	2\z = \frac{\D K}{Kd},
\end{align}
or equivalently by \eqref{Ds-B2}
\begin{align}\label{eqn:dKtophi1}
	s_\CB = 2L \sqrt{\frac{|\D L|}{L}} = 2 L \sqrt{\frac{|\D K|}{Kd} .}
\end{align}
Notice that we added absolute values since these equations were derived assuming $\D L > 0$ (see Footnote~\ref{fn:DL}).

We remark that \eqref{eqn:dKtophi} was first derived in \cite{Verlinde:2019xfb, Verlinde:2019ade} using the fact that we can write $\zeta$ in terms of an effective Newtonian potential
\begin{equation}\label{zeta-mass}
	\TBHPhi = \frac{8 \pi G_N M}{d \Omega_d L^{d-1}}  , 
\end{equation}
where the ``mass'' $M$ is sourced by modular Hamiltonian fluctuations
\begin{equation}\label{eqn:MinK}
	M = \frac{\D K}{4 \pi L}  .
\end{equation}
Substituting this into \eqref{zeta-mass}, we arrive at \eqref{eqn:dKtophi}. Furthermore, the analysis of \cite{Verlinde:2019xfb,Verlinde:2019ade} is done in the ``canonical ensemble'' where $\la (\D K)^2 \ra_{n=1} \not= 0$. However, as was shown in Section~\ref{subsec:LT}, we can by \eqref{Dn-1} equivalently describe fluctuations in the horizon after performing a Legendre transform as
\begin{equation}\label{eqn:dntophi}
	2 \TBHPhi = - \frac{\Delta n}{d} \, .
\end{equation}

We conclude this subsection by remarking that while \cite{Verlinde:2019xfb, Verlinde:2019ade} implicitly assumed $\ka = \frac{1}{2L}$, we can reinstate $\ka$ by taking $2L \to \frac{1}{\ka}$ in \eqref{eqn:MinK}, so that $M = \frac{\ka \D K}{2\pi}$. In that case, we would find
\begin{align}\label{eqn:dLalt}
	\frac{\D L}{L} = 2\ka L \frac{\D K}{K d},
\end{align}
which appears to be a less natural relation between $L$ and $K$. Thus, even though $\ka$ ultimately drops out of any physical quantities discussed in the next subsection, we will henceforth fix $\ka = \frac{1}{2L}$.

\subsection{Fluctuations in a Photon Trajectory}\label{subsec:photon-fluc}

A causal diamond can be naturally viewed as an interferometer setup, where a laser is emitted from the ``beamsplitter'' at the origin in all directions, is reflected on a mirror, and is finally observed when it again reaches the beamsplitter. In this setup, the phase shifts between two different directions or a local reference can in principle be measured. Therefore, we will consider such a phase shift as a physical observable, although our analysis here is simply meant to illustrate an important scaling principle in our setup, first proposed in \cite{Verlinde:2019xfb}, and is not at a level where it is directly applicable to a real experiment. Nevertheless, as in \cite{Lee:2024oxo}, we will extract a phase shift for the interferometer setup from the differences between the two worldlines in the perturbed and unperturbed geometries.

Our approach is as follows. First, we will compute the phase difference that arises when a photon travels along a stretched horizon rather than the causal horizon. From the perspective of the perturbed geometry, this photon has an effective mass, and the description formally resembles that of the propagation of electromagnetic waves through an optical medium (e.g., see \cite{Ebadi:2023gne}). A similar computation of the photon phase shift was done recently in \cite{Lee:2024oxo}, which we show in Appendix~\ref{app:photon} to reproduce our results in this subsection.

Given a photon whose trajectory is along a stretched horizon $\CH_s$ in the causal diamond, its accumulated phase shift is given by
\begin{align}\label{eqn:dpsialongstretch}
\begin{split}
	\D\psi = \int_{\tau_-}^{\tau_+} \dt\tau \, \w(\tau),
\end{split}
\end{align}
where $\tau$ is the proper time along $\CH_s$, and $\w(\tau) = U^{\tt}$, with $U^\mu$ being the four-velocity along $\CH_s$, is the energy with respect to the Minkowski time $\tt$. Explicitly, $U^\mu$ in Minkowski coordinates is given by
\begin{align}\label{4-velocity}
\begin{split}
	U^\mu\p_\mu = \lambda \frac{\dt x^\mu}{\dt \tau}\p_\mu = \lambda \bigg( \frac{\dt \tt}{\dt \tau} \p_\tt + \frac{\dt \tr}{\dt \tau}\p_\tr \bigg),
\end{split}
\end{align}
where $\lambda$ is a normalization constant. To compute $U^\mu$, we need to determine the induced metric on $\CH_s$. Recalling that the stretched horizon given by $r = \D L$ is parametrized using Minkowski coordinates by \eqref{eqn:Rindlerrint}, we obtain
\begin{align}\label{t-r-mink}
\begin{split}
	2\tt\,\dt\tt = - 2 \sqrt{\tt^2 + s_\CB^2} \,\dt\tr.
\end{split}
\end{align}
Substituting this into \eqref{4-velocity}, we get
\begin{align}\label{eqn:u}
\begin{split}
	U^\mu \p_\mu &= \lambda \frac{\dt \tt}{\dt\tau} \Bigg(\p_\tt - \frac{\tt}{\sqrt{\tt^2 + s_\CB^2} } \p_\tr \Bigg) .
\end{split}
\end{align}
Next, to determine $\frac{\dt x^\mu}{\dt\tau}$, note that the proper time along $\CH_s$ is given by
\begin{align}
\begin{split}
	-\dt \tau^2\big|_{\CH_s} = \dt s^2\big|_{\CH_s} = -\Bigg( 1 - \frac{\tt^2}{\tt^2 + s_\CB^2} \Bigg) \dt \tt^2,
\end{split}
\end{align}
which implies
\begin{align}\label{prop-time}
\begin{split}
	\frac{\dt \tt}{\dt \tau}  = \sqrt{1 + \frac{\tt^2}{s_\CB^2 }} .
\end{split}
\end{align}
Substituting this back into \eqref{eqn:u}, we see that the four velocity is given by
\begin{align}\label{eqn:u2}
\begin{split}
	U^\mu\p_\mu &=  \lambda \Bigg( \sqrt{1 + \frac{\tt^2}{s_\CB^2}} \p_\tt - \frac{\tt}{s_\CB}\p_\tr \Bigg) .
\end{split}
\end{align}

We would now like to fix the normalization constant $\lambda$. Let $\w_0$ be the laser frequency at the beamsplitter. As $U^\tt$ is associated to the energy, we normalize the four-velocity so that $U^\tt = \w_0$ at $\tt = \tt_-$, the past tip of $\CH_s$. This implies
\begin{align}\label{normalization}
\begin{split}
	\lambda = \frac{\w_0}{\sqrt{1 + \frac{\tt_-^2}{s_\CB^2} } } ,
\end{split}
\end{align}
and so the effective mass is given by
\begin{align}\label{eqn:effmass0}
\begin{split}
	m_\eff^2 &= - U^\mu U_\mu = \lambda =  \frac{\w_0^2}{ 1 + \frac{\tt_-^2}{s_\CB^2} } .
\end{split}
\end{align}
Recalling \eqref{t-dif}, we have
\begin{align}\label{eqn:effmass}
\begin{split}
	m_\eff &= \frac{\w_0 s_\CB}{L} .
\end{split}
\end{align}

We can now use the effective mass to compute the phase shift. Noting that $\lambda = m_\eff$ from \eqref{eqn:effmass0}, we have
\begin{align}
\begin{split}
	\D \psi &= \int_{\tau_-}^{\tau_+} \dt\tau\, U^\tt =  m_\eff  \int_{\tau_-}^{\tau_+} \dt\tau\, \frac{\dt\tt}{\dt\tau} = m_\eff (t_+ - t_-) .
\end{split}
\end{align}
Substituting \eqref{t-dif} and \eqref{eqn:effmass} into the above equation, we find
\begin{align}\label{phase-shift-fin}
\begin{split}
	\D\psi &= 2\w_0s_\CB  + \CO(s_\CB^2).
\end{split}
\end{align}
Thus, we see to leading order, $\D\psi$ scales as $s_\CB \sim (\D L)^{\frac{1}{2}}$. We can also write this in terms of $\D K$ by substituting in $s_\CB$, in which case the phase shift is to leading order
\begin{align}\label{phase-shift-fin2}
\begin{split}
	\D\psi = 4 L \w_0 \sqrt{\frac{|\D K|}{K d} } ,
\end{split}
\end{align}
where we used \eqref{eqn:dKtophi1}. Dividing both sides by $2 \w_0 L$ and squaring, we obtain the fluctuation in the strain $h$ during one photon traversal:
\begin{align}
\begin{split}
	h^2 \equiv \left( \frac{s_{\cal B}}{L}\right)^2 = \frac{4|\D K|}{K d}  . 
\end{split}
\end{align} 
This result is consistent with other calculations \cite{Verlinde:2019xfb, Verlinde:2019ade, Gukov:2022oed}, and improves on them by computing an explicitly gauge-invariant quantity associated to an observable of a photon traversing a causal diamond.

\section{Discussion} \label{discussion}
		
We have studied, utilizing thermodynamics, the size of modular fluctuations, or equivalently area fluctuations, of causal diamonds in Minkowski space. We were able to carry out this thermodynamic analysis by first deriving an on-shell action associated to a finite causal diamond, and identifying it in the semiclassical limit with the logarithm of the partition function $-\log\CZ$. By using the replica trick, we were then able to compute both the mean and variance of modular fluctuations. The modular fluctuations give rise to changes in the geometry, which can be observed as a photon phase shift arising from lasers traversing the causal diamond boundaries.

Furthermore, we have also interpreted variations in the replica index as variations in the Rindler observers associated to the causal horizons. As long as our causal diamond is sufficiently large such that we may treat a family of Rindler observers in the planar Rindler limit, we can view such variations as fluctuations in their Unruh temperatures. This result also validates the usual analytic continuation of the replica index to non-integer values when performing the replica trick, as the value of the replica index simply parametrizes the acceleration of a Rindler observer, which certainly needs not be integer-valued.

There are still many interesting questions that remain. In particular, our analysis throughout relied on thermodynamic identities, but the microscopic dynamics that give rise to such identities remain mysterious. We do not have a complete understanding of quantum metric fluctuations that could give rise to such thermodynamic behavior. Nevertheless, steps towards understanding the microscopic behavior of quantum gravity near a causal horizon utilizing a fluid description were recently proposed in \cite{Zhang:2023mkf, Bak:2024kzk}.  A deeper understanding of these steps, employing a Schwinger-Keldysh (or Feynman-Vernon) formalism, is already suggested by the regularization of the on-shell action deployed here, and will be explored in future work.

Another important question is to better understand the reduced density matrix associated to a causal diamond, and how this density matrix is derived from the vacuum structure of gauge theory and gravity. The recent advances in the understanding of vacua in gauge theories and gravity (e.g., see \cite{He:2014laa, He:2020ifr, Kapec:2022hih, He:2023bvv}) allow for nontrivial entanglement among an infinity of possible vacuum states. It would be illuminating to construct such a density matrix for our empty causal diamond such that the various thermodynamic results we obtained in this paper can be derived directly from the state. This will allow us to go beyond computing expectation values and provide instead operator identities, and is a topic that we are actively exploring.

%
%
	
\section*{Acknowledgments}

We would like to thank Ning Bao, Mathew Bub, Clifford Cheung, Luca Ciambelli, Bartlomiej Czech, Laurent Freidel, Marc Klinger, Prahar Mitra, Hirosi Ooguri, and Julian Sonner for helpful discussions. K.F., T.H., and K.Z. are supported by the Heising-Simons Foundation “Observational Signatures of Quantum Gravity” collaboration grant 2021-2817, the U.S. Department of Energy, Office of Science, Office of High Energy Physics, under Award No. DE-SC0011632, and the Walter Burke Institute for Theoretical Physics. T.H. would like to thank the Galileo Galilei Institute for Theoretical Physics for their hospitality and the INFN for partial support during the completion of this work. K.Z. is also supported by a Simons Investigator award.

\appendix

\section{Coordinate Systems}\label{app:coordinates}

For convenience and to fix conventions, we collect here the various relevant coordinate systems describing a spherically symmetric causal diamond in Minkowski spacetime. We will summarize their relations to each other and, in particular, to the standard Minkowski metric in polar coordinates given by
\begin{align}\label{mink-coord}
\begin{split}
	\dt s^2 = -\dt \tt^2 + \dt \tr^2 + \tr^2\,\dt\O_d^2 ,
\end{split}
\end{align}
where the causal diamond is defined by the region $|\tt|  \leq L - \tr$. In spherical lightcone coordinates, we can rewrite the above metric as
\begin{align}\label{eqn:lc}
\begin{split}
	\tu = \tt - \tr , \qquad \tv = \tt + \tr, \qquad \dt s^2 = -\dt \tu\,\dt\tv + \tr^2 \, \dt\O_d^2 .
\end{split}
\end{align}

\subsection{Spherical Rindler Coordinates}

The relation between spherical Rindler coordinates\footnote{Note that we use the term spherical Rindler coordinates to describe the interior complement of the region accessible to outwardly accelerating radial observers.} $(T, \eta)$ and the light-cone coordinates in \eqref{eqn:lc} is given by
\begin{equation}\label{uv-to-etaT}
	\tilde{u} = - L + \eta e^{\kappa T} , \qquad \tilde{v} = L - \eta e^{-\kappa T}   .
\end{equation}
Note that this also implies
\begin{align}\label{rindler-r}
	\tr = \frac{1}{2}(\tv - \tu) = L - \eta \cosh(\ka T).
\end{align}
Substituting \eqref{uv-to-etaT} into \eqref{eqn:lc}, we get the spherical Rindler metric
\begin{equation}\label{eqn:rindler}
	\dt s^2 = -(\kappa \eta)^2 \, \dt T^2 + \dt \eta^2 + \tilde{r}(T,\eta)^2 \,\dt \Omega_d^2  ,
\end{equation}
with $\tr(T,\eta)$ given in \eqref{rindler-r}.

In Euclidean signature where we take $T = - i T_E$, the Euclidean metric becomes
\begin{align}
\begin{split}
	\dt s_E^2 &= (\ka \eta)^2 \,\dt T_E^2 + \dt\eta^2 + \tr_E(T_E,\eta)^2\,\dt\O_d^2 \\
	\tr_E(T_E,\eta) &= L - \eta \cos(\ka T_E).
\end{split}
\end{align}
Note that the $(T_E,\eta)$ part of the metric is simply polar coordinates, and therefore is smooth only if we have the periodic identification
\begin{align}
	T_E \sim T_E + \frac{2\pi}{\ka}.
\end{align}
It is obvious $\tr_E(T_E,\eta)$ remains unchanged under this identification as well. 

Unlike planar Rindler coordinates describing acceleration in a fixed Cartesian direction (or parallel observers), the time translation vector $\p_T$ in spherical Rindler coordinates does not generate an isometry due to the explicit time dependence in \eqref{eqn:rindler}. Explicitly, we observe
\begin{align}
\begin{split}
	\CL_{\p_T} \dt s^2 &= \p_T (\dt s^2) = - 2 \ka\eta \sinh(\ka T) ( L - \eta\cosh(\ka T))\,\dt\O_d^2.
\end{split}
\end{align}
We refer the reader to \cite{Balasubramanian:2013rqa} for a discussion on some subtleties related to this point.

\subsection{Topological Black Hole Coordinates}

Next, in order to go from spherical Rindler coordinates $(T,\eta)$ to topological black hole (TBH) coordinates $(T, R)$, we use the coordinate map
\begin{equation}
	\eta^2 = \frac{2}{\kappa} f(R) , \qquad f(R) = L (1 + 2\zeta)  -R ,
\end{equation}
where $\z$ is a spacetime constant introduced to match the conventions of \cite{Verlinde:2019xfb}. Substituting this into \eqref{rindler-r} and \eqref{eqn:rindler}, we get
\begin{align}\label{eqn:tbh-met}
\begin{split}
	\dt s^2 &= - 2  \kappa f(R) \,\dt T^2 + \frac{\dt R^2}{2  \kappa f(R)} + \tilde{r}(T,R)^2 \,\dt\Omega_d^2 \\
	\tr(T,R) &= L - \sqrt{\frac{2f(R)}{\ka}}\cosh(\ka T).
\end{split}
\end{align}
We can also choose to center the coordinates on the horizon with the coordinate map  
\begin{equation}
	r = f(R) , \qquad t = T  ,
\end{equation}
in which case we have
\begin{align}\label{eqn:TBHcentered}
\begin{split}
	\dt s^2 &=  - 2 \kappa r \,\dt t^2 + \frac{\dt r^2}{2 \kappa r} + \tilde{r}(t,r)^2 \, \dt\Omega_d^2  \\
	\tr(t,r) &= L - \sqrt{\frac{2r}{\ka}}\cosh(\ka t),
\end{split}
\end{align}
where $r$ should not be confused with the usual areal radius of the transverse sphere, which we denoted $\tilde{r}$. Now, we define a natural ``tortoise'' coordinate, and its associated light-cone coordinates, to be
\begin{equation}\label{eqn:TBHcenteredtoGN}
	u = t + r^*, \qquad v = t - r^* , \qquad r^* \equiv  \frac{1}{2 \kappa} \log 2 \kappa r -\frac{\alpha}{\kappa}  ,
\end{equation}
where the choice of integration constant (involving $\alpha$) in the definition of $r^*$ is informed by comparing it to \eqref{eqn:metric}.

\subsection{Gaussian Null Coordinates}

We now map the metric \eqref{eqn:TBHcentered} to Gaussian null coordinates. Using \eqref{eqn:TBHcenteredtoGN} to write $t = u - r^*$, the metric \eqref{eqn:TBHcentered} in Gaussian null coordinates is
\begin{equation}\label{eqn:app:inGN}
	\dt s^2 =  - 2 \kappa r\, \dt u^2 + 2\, \dt u\, \dt r + \bigg( L - \frac{1}{2 \kappa}e^{\kappa u+ \alpha} - re^{-\kappa u-\alpha} \bigg)^2 \dt\Omega_d^2 ,
\end{equation}
which is precisely \eqref{eqn:metric}.

Equivalently we could have used advanced coordinates $t= v + r^*$, in which case the metric \eqref{eqn:TBHcentered} becomes
\begin{equation}
	\dt s^2 =  - 2 \kappa r\, \dt v^2 - 2 \,\dt v \dt r + \bigg( L - \frac{1}{2 \kappa}e^{-\kappa v+ \alpha}- re^{\kappa v-\alpha} \bigg)^2 \, \dt \Omega_d^2 .
\end{equation}
As noted in the main text (also, see \cite{Bub:2024nan}), the metric \eqref{eqn:metric} is only a particular way to write flat metrics in a more general class of metrics in Gaussian null coordinates. 

\subsection{Conformal Killing Coordinates}\label{app:confcoordinates}

While a spherically symmetric causal diamond in Minkowski spacetime is not preserved by the flow of a Killing time, it is conserved by that of a conformal Killing time. Therefore, while not an equilibrium state, such causal diamonds are still ``conformally stationary'' \cite{Jacobson:2018ahi}. In this subsection, we will construct coordinate adapted to such a conformal Killing time and discuss their relation to the other coordinate systems we have examined above. A more in-depth analysis of conformal Killing flows was also performed recently in \cite{Caminiti:2025hjq}.

First, to construct the conformal Killing vector (CKV) $\z$, we recall that it is defined by the fact the metric is allowed to rescale under the flow generated by $\z$, so that
\begin{align}\label{ckv-def}
	\CL_\z g_{\mu\nu} = \nabla_{\mu} \z_\nu + \nabla_\nu \z_\mu = \alpha g_{\mu\nu},
\end{align}
where $\alpha$ is some spacetime function. This implies the conformal Killing equation
\begin{align}\label{ckv-fin}
	\nabla_\mu \z_\nu + \nabla_\nu \z_\mu = \frac{2}{d+2}(\nabla \cdot \z ) g_{\mu\nu}.
\end{align}
To determine what is $\z$ for our spherically symmetric causal diamond in Minkowski coordinates using \eqref{mink-coord}, note that $\z$ must vanish at the tips of the diamond, namely at $(\tt,\tr) = (\pm L, 0)$ and $(\tt,\tr) = (0, L)$, since the flow needs to keep the boundaries invariant. Using \eqref{ckv-fin} with the above boundary conditions, we arrive at
\begin{align}\label{eqn:app:CKV}
	\zeta = \mu\Big[ ( L^2 - \tt^2 - \tr^2)\p_\tt - 2\tt\tr\p_\tr \Big] ,
\end{align}
where $\mu$ is an arbitrary normalization constant.\footnote{In the limit $L \to \infty$, we have $\z \to \mu L^2\p_\tt$. In this case, the CKV becomes a true Killing vector, and we can choose $\mu = \frac{1}{L^2}$ so that $\z^2 = -1$ is normalized to unity. For finite $L$ however, it is less clear what is a preferred choice of normalization. \label{fn:norm} } It is straightforward to check that on the boundaries of the causal diamond we have $\z^2 = 0$, implying that the causal diamond boundary is precisely the conformal Killing horizon. On a conformal Killing horizon $\CH$, the surface gravity is defined via the equation
\begin{align}
	\nabla_\mu(\z^2)\big|_\CH = -2\ka \z_\mu .
\end{align}
Evaluating this for both $\CH^\pm$, defined by $\tt = \pm(L - \tr)$, we obtain
\begin{align}\label{surface-grav}
\begin{split}
	\nabla_\mu(\z^2)\big|_{\CH^\pm} = -2\ka_\pm \z_\mu \quad\implies\quad \ka_\pm = \pm 2 L\mu,
\end{split}
\end{align}
where the minus sign for the past horizon is due to the fact the normal vector of the past horizon is past-directed rather than future-directed. 

Following Appendix B of \cite{Jacobson:2018ahi}, we would now like to construct a coordinate system adapted to the flow of $\z$. Let $\sigma$ denote the conformal Killing time, and we choose it such that the $\s=0$ and $t=0$ hypersurfaces coincide. The coordinate system by construction should have $\z = \p_\s$ be the conformal Killing vector, which implies that we want the past and future horizons be located infinitely far away at $\s = \pm\infty$, respectively. Furthermore, we will choose a spherically symmetric coordinate $x$ on the constant $\s$ slices, so that $x=0$ coincides with $r=0$, and $x \to \infty$ as we approach the horizons $\CH^\pm$. It follows the metric takes the form
\begin{align}\label{eqn:met-conf}
	\dt s^2 &= C(\s,x)^2 \big( - \dt \s^2 + \dt x^2 \big) + \tr(\s,x)^2\,\dt\O_d^2  .
\end{align}
To ensure that $\z = \p_\s$ is a CKV,\footnote{We define $\z$ to be the vector generating conformal Killing time, so its normalization is set to be $1$.} we require it to satisfy the conformal Killing equation \eqref{ckv-fin}. This implies $\tr(\s,x) = C(\s,x)\rho(x)$, where $\rho(x)$ is a function of $x$ only, so that the metric becomes
\begin{align}\label{eqn:metricconformal}
	\dt s^2 &= C(\s,x)^2 \big( - \dt\s^2 + \dt x^2 + \rho(x)^2\,\dt\O_d^2 \big) .
\end{align}
Furthermore, \eqref{surface-grav} implies the surface gravity $\ka_\pm$ on $\CH^\pm$ satisfies
\begin{align}\label{consistency}
\begin{split}
	\ka_\pm = \lim_{\s \to \pm\infty} - \frac{\p_\s C}{C}, \qquad \lim_{\s \to \pm \infty} \p_x C = 0.
\end{split}
\end{align}

Next, let us determine the relation between the lightcone coordinates $(\tu,\tv)$ given in \eqref{eqn:lc} and the conformal lightcone coordinates
\begin{align}
	\bar u = \s - x , \qquad \bar v = \s + x.
\end{align}
In these coordinates, the CKV is given by
\begin{align}\label{lc-ckv}
	&\z = \p_\s =   \p_{\bar u} + \p_{\bar v}  .
\end{align}
On the other hand, note that from \eqref{eqn:app:CKV}, the CKV in standard lightcone coordinates is given to be
\begin{align}\label{conformal-lc-ckv}
\begin{split}
	\z &=  \mu\Big[ ( L^2 - \tu^2 )  \p_\tu + ( L^2 - \tv^2 ) \p_\tv  \Big] .
\end{split}
\end{align}
It is clear from the form of the CKV that we expect $\bu \equiv \bu(\tu)$ and $\bv \equiv \bv(\tv)$. Thus, matching the components of the CKV between \eqref{lc-ckv} and \eqref{conformal-lc-ckv}, we get the relations
\begin{align}
\begin{split}
	&\tu = L\tanh \big(\mu L \bu \big), \qquad \tv = L\tanh  \big( \mu L \bv \big), 
\end{split}
\end{align}
so that
\begin{align}
\begin{split}
	\tt &= \frac{L}{2} \Big[ \tanh\big( \mu L (\s-x) \big) + \tanh \big( \mu L (\s+x) \big) \Big]  \\
	\tr &= \frac{L}{2} \Big[ \tanh \big( \mu L (\s+x) \big) - \tanh \big( \mu L (\s-x) \big) \Big] .
\end{split}
\end{align}
Substituting this into \eqref{mink-coord}, we arrive at
\begin{align}\label{ckv-metric}
\begin{split}
	\dt s^2 &= \bigg( \frac{2\mu L^2}{\cosh ( 2\mu  L\s ) + \cosh ( 2\mu Lx )} \bigg)^2 ( -\dt \s^2 + \dt x^2) + \tr(\s,x)^2 \dt\O_d^2 ,
\end{split}
\end{align}
thereby allowing us to deduce
\begin{align}\label{eqnmetriconformal2}
\begin{split}
	C(\s,x) &=  \frac{2\mu L^2}{\cosh( 2\mu L\s ) + \cosh ( 2\mu Lx )}  \\
	\rho(x) &= \frac{\tr(\s,x)}{C(\s,x)}  = \frac{1}{2L\mu} \sinh (2L\mu x ).
\end{split}
\end{align}
We observe that $C(\s,x)$ satisfies the condition $\lim_{\s \to \pm \infty} \p_x C = 0$, as required by \eqref{consistency}. Furthermore, the surface gravity by \eqref{consistency} is
\begin{align}
\begin{split}
	\ka_\pm = \pm 2 L \mu,
\end{split}
\end{align}
matching \eqref{surface-grav} as required.

Finally, we can use the conformal Killing metric \eqref{ckv-metric} to compute the following quantities. At constant conformal time $\s = \s_0$, the (radially outward) proper distance between the points $x=0$ on the beamsplitter and $x \to \infty$ on the horizon is given by
\begin{align}
	\int_0^\infty \dt x\, \frac{2\mu L^2}{\cosh( 2\mu L \s_0 ) + \cosh (2\mu Lx )}  = \frac{2L\mu\s_0}{\sinh(2\mu L \s_0)} .
\end{align}
Similarly, the proper time an accelerated observer at $x = x_0$ needs to traverse the diamond is given by
\begin{align}
\begin{split}
	\int_{-\infty}^{\infty} \dt \s \, \frac{2\mu L^2}{\cosh( 2\mu L \s ) + \cosh(2\mu L x_0 )} = \frac{4L \mu x_0}{\sinh( 2\mu L x_0)} .
\end{split}
\end{align}

\section{On-shell Action Regularization}\label{app:onshell}

In this appendix, we will regularize the on-shell action using two different coordinate systems. In Appendix~\ref{app:gn-reg}, we will perform the regularization in Gaussian null coordinates. In Appendix~\ref{app:tbh-reg}, we will derive the on-shell action, as well as explain how to regularize it, in topological black hole coordinates.

\subsection{Gaussian Null Coordinates with $r \not= 0$}\label{app:gn-reg}

In this subsection, we will compute the on-shell action slightly off of the causal horizon and instead assume our boundary is a stretched horizon at $r=r_0$. Our starting point is \eqref{eqn:thetastart2rneq0} with boundary condition $L = L_\CB$ (given in \eqref{fixed-area}), which we recall here for convenience:\footnote{The $+$ superscript on $\widetilde\Th_{\CH_s}^+$ reminds us we are regularizing the symplectic potential associated to the future horizon $\CH^+$.}
\begin{align}\label{pre-symp-r0app}
\begin{split}
	\widetilde\Th_{\CH_s}^+ &= \frac{\O_d}{8\pi G_N} \int_{\CH_s^\reg} \dt u \bigg[ \Phi^d \bigg( \delta \ka + d \frac{\p_u\delta\Phi}{\Phi} \bigg) + r_0 d \Phi^{d-1}\big( \d \ka \p_r\Phi + 2\ka \p_r\d\Phi \big)  \bigg]\bigg|_{r=r_0} \\
	\Phi &= L_\CB - \frac{1}{2\ka} e^{\ka u + \a} - r e^{-\ka u - \a} ,
\end{split}
\end{align}
where $\CH_s^\reg$ indicates we are only integrating over the part of the stretched horizon that limits to $\CH^+$. In Section~\ref{sec:onshell}, we immediately took $r=0$ and therefore dropped the second term in the above integral. We would now like to reconsider the calculation at $r=r_0$ and take the $r_0 \to 0$ limit more explicitly.

In this case, we have from \eqref{eqn:mu}
\begin{align}\label{mu-r0}
\begin{split}
	\mu(u,r_0) &= \ka u + \a + \frac{1}{2} \log \big( 1 - 2\ka r_0 e^{-2(\ka u + \a)} \big) .
\end{split}	
\end{align}
Notice that if we set $r_0 = 0$, as we had done in \eqref{mu-prop}, we would simply get $\mu = \ka u + \a$. On the other hand, it is possible to take $r_0 \to 0$ and $u \to -\infty$ such that the last term in \eqref{mu-r0} does not vanish. Following the calculations done in Section~\ref{sec:onshell}, we have
\begin{align}\label{mu-prop-app}
\begin{split}
	\p_u \mu(u,r_0) &= \ka + \frac{2\ka^2 r_0 e^{-2(\ka u + \a)}}{1-2\ka r_0 e^{-2(\ka u + \a)}} \\
	\d \mu(u,r_0) &= \d (\ka u + \a) - \frac{\d \big( \ka r_0 e^{-2 (\ka u + \a)} \big)}{1-2\ka r_0 e^{-2(\ka u + \a)}}  \\
	&= \frac{\d(\ka u + \a) - \d(\ka r_0) e^{-2(\ka u + \a)}}{1- 2\ka r_0 e^{-2(\ka u + \a)}}  .
\end{split}
\end{align}
Next, we observe
\begin{align}\label{Phi-app}
\begin{split}
	\p_u\Phi(u,r_0) &= -\frac{1}{2} e^{\ka u + \a} + \ka r_0 e^{-\ka u - \a} \\
	\p_u \d \Phi(u,r_0) &= - \frac{1}{2} e^{\ka u + \a} \d(\ka u + \a ) + \d\big( \ka r_0 e^{-\ka u - \a} \big),
\end{split}
\end{align}
from which we conclude
\begin{align}\label{mu-prop-app2}
\begin{split}
	& \frac{\p_u\d\Phi(u,r_0)}{\p_u\Phi(u,r_0)} = \frac{-\frac{1}{2}e^{\ka u +\a} \d(\ka u + \a) + \d\big( \ka r_0 e^{-\ka u -\a } \big) }{-\frac{1}{2} e^{\ka u + \a} + \ka r_0 e^{-\ka u - \a} } \\
	\implies\quad & \d(\ka u + \a) = \big( 1 - 2\ka r_0 e^{-2(\ka u + \a)} \big) \frac{\p_u\d\Phi(u,r_0)}{\p_u\Phi(u,r_0)} + 2e^{-\ka u - \a}\d\big(\ka r_0 e^{-\ka u -\a}\big)  .
\end{split}
\end{align}
Substituting this into the last line in \eqref{mu-prop-app}, we get
\begin{align}\label{eqn:mucorr}
\begin{split}
	\d\mu(u,r_0) &= \frac{\p_u\d\Phi(u,r_0)}{\p_u\Phi(u,r_0)} + \frac{2e^{-\ka u - \a}\d\big(\ka r_0 e^{-\ka u -\a}\big) - \d(\ka r_0) e^{-2(\ka u + \a)}}{1 - 2\ka r_0 e^{-2(\ka u + \a)}} \\
	&= \frac{\p_u\d\Phi(u,r_0)}{\p_u\Phi(u,r_0)} + \frac{\d\big( \ka r_0 e^{-2(\ka u + \a)} \big) }{1 - 2\ka r_0 e^{-2(\ka u + \a)}} .
\end{split}
\end{align}
Substituting the first equality in \eqref{mu-prop-app} and \eqref{eqn:mucorr} into \eqref{pre-symp-r0app}, we get
\begin{align}\label{eqn:thetastart3rneq0}
\begin{split}
	\widetilde\Th_{\CH_s}^+ &= \frac{\O_d}{8\pi G_N} \int_{u_\reg}^{u_+}  \dt u \Bigg[ \big( \Phi^d \p_u \d \mu + d\Phi^{d-1}\p_u\Phi \d\mu \big) + r_0 d \Phi^{d-1}\big( \d \ka \p_r\Phi + 2\ka \p_r\d\Phi \big)   \\
	&\qquad - \Phi^d \d \Bigg( \frac{2\ka^2 r_0 e^{-2(\ka u + \a)}}{1-2\ka r_0 e^{-2(\ka u + \a)}} \Bigg) - \p_u(\Phi^d) \frac{\d\big( \ka r_0 e^{-2(\ka u+\a)} \big)}{1-2\ka r_0 e^{-2(\ka u + \a)}} \Bigg]\Bigg|_{r=r_0} \\ 
	&= \frac{\O_d}{8\pi G_N} \int_{u_\reg}^{u_+} \dt u \Bigg[ \p_u\big( \Phi^d \d\mu \big) + r_0 d \Phi^{d-1}\big( \d \ka \p_r\Phi + 2\ka \p_r\d\Phi \big)  \\
	&\qquad   - \p_u \Bigg(\Phi^d \frac{\d\big( \ka r_0 e^{-2(\ka u+\a)} \big)}{1-2\ka r_0 e^{-2(\ka u + \a)}} \Bigg) + 2\Phi^d  \p_u \Bigg( \frac{\d \big( \ka r_0 e^{-2(\ka u + \a)}\big)}{1-2\ka r_0 e^{-2(\ka u + \a)}} \Bigg) \Bigg]\Bigg|_{r=r_0}
\end{split}
\end{align}
where we wrote the integration limits explicitly to be $u_\reg$ and $u_+$, with the understanding we may take $u_\reg = u_\mid$ eventually. Notice that in the final expression, except for the first term, the other three terms were not present in our previous analysis at $r=0$, and correct our result \eqref{eqn:thetastart3}.

Let us now focus on each of the terms in the integrand on the right-hand side of \eqref{eqn:thetastart3rneq0}. We will henceforth assume that $r_0$ and $u$ are phase space constants and hence do not vary (i.e., $\d u = \d r_0 = 0$). With this assumption, the first term is essentially the regularized version of \eqref{eqn:thetastart3}, in that
\begin{align}\label{I1}
\begin{split}
	\SI_1 &\equiv \frac{\O_d}{8\pi G_N} \int_{u_\reg}^{u_+} \dt u\,\p_u(\Phi^d\d \mu) \\
	&= -\frac{\O_d}{8\pi G_N} \Phi(u_\reg,r_0)^d \d\mu(u_\reg,r_0) \\
	&= -\frac{\O_d}{8\pi G_N} \bigg( L_\CB - \frac{1}{2\ka}e^{\ka u_\reg+\a} - r_0 e^{-\ka u_\reg-\a} \bigg)^d  \frac{\d\ka u_\reg + \d\a - r_0 \d\ka e^{-2(\ka u_\reg+\a)}}{1-2\ka r_0 e^{-2(\ka u_\reg+\a)}} 
\end{split}
\end{align}
where in the second equality we used the fact $\Phi(u_+,r_0) = 0$ by construction,  and in the last equality we used \eqref{pre-symp-r0app} and \eqref{mu-prop-app}. Let us now assume the midpoint of the stretched horizon is where we want $u_\reg$ to be, so that recalling \eqref{eqn:umax}, we have
\begin{align}\label{u-reg-assume}
\begin{split}
	u_\reg \equiv u_\mid = - \frac{\a}{\ka} + \frac{1}{2\ka}\log(2\ka r_0) + \e \quad\implies\quad e^{\ka u_\reg + \a}  = e^{\ka \e}\sqrt{2\ka r_0},
\end{split}
\end{align}
where $\e$ is a regularization parameter in that we will eventually take $\e \to 0$. Substituting this into \eqref{I1}, we get upon taking the $r_0 \to 0$ limit
\begin{align}\label{I1-fin}
\begin{split}
	\lim_{r_0 \to 0} \SI_1 &= \lim_{r_0 \to 0} \frac{1}{2\ka} \frac{\O_d}{8\pi G_N} L_\CB^d  \frac{(e^{-2\ka\e} - 2\ka u_\reg) \d\ka  - 2\ka \d\a  }{1- e^{-2\ka\e}} .
\end{split}
\end{align}
Note we do not take the $\e\to 0$ limit yet, even in the numerator, as we will have to keep track of the finite terms, i.e., the terms subleading in a small $\e$ expansion, in addition to the divergent terms. Furthermore, we keep the $r_0 \to 0$ limit explicit on the right-hand side since $u_\reg$ depends on $r_0$.

Next, the second term of the integrand in \eqref{eqn:thetastart3rneq0} is 
\begin{align}\label{I2-int}
\begin{split}
	\SI_2 &\equiv \frac{\O_dr_0 d}{8\pi G_N} \int_{u_\reg}^{u_+} \dt u \,   \Phi^{d-1}\big( \d \ka \p_r\Phi + 2\ka \p_r\d\Phi \big) \\
	&= -\frac{\O_d \ka r_0 d}{4\pi G_N} \int_{u_\reg}^{u_+} \dt u \,   \Phi^{d-1}\bigg( \frac{\d \ka}{2\ka } - \d(\ka u + \a) \bigg) e^{-\ka u - \a}.
\end{split}
\end{align}
Now, notice for any $k \geq 1$, we can perform the series expansion
\begin{align}\label{Phi-series}
\begin{split}
	\Phi(u,r_0)^k = \sum_{\substack{k_1 + k_2 + k_3 = k \\ k_1,k_2,k_3 \geq 0}} (-1)^{k_2+k_3} \binom{k}{k_1,k_2,k_3} L_\CB^{k_1} \bigg( \frac{1}{2\ka} e^{\ka u + \a} \bigg)^{k_2} (r_0 e^{-\ka u - \a})^{k_3} .
\end{split}
\end{align}
It follows we can rewrite the integral in \eqref{I2-int} as a sum of integrals proportional to (keeping track of only $r_0$ and $u$ dependences)
\begin{align}\label{integral-forms}
\begin{split}
	J_1^{\lambda_1,\lambda_2}(r_0) &\equiv r_0^{\lambda_1} \int_{u_\reg}^{u_+} \dt u\, e^{\lambda_2(\ka u + \a)} \\
	&= \frac{r_0^{\lambda_1}}{\lambda_2\ka} \big( e^{\lambda_2(\ka u_+ + \a)}  -  e^{\lambda_2(\ka u_\reg + \a)}  \big) \\
	J_2^{\lambda_1,\lambda_2}(r_0) &\equiv r_0^{\lambda_1} \int_{u_\reg}^{u_+} \dt u\, u e^{\lambda_2(\ka u + \a)} \\
	&= \frac{r_0^{\lambda_1}}{\lambda_2^2\ka^2} \Big[ \big( 1- \lambda_2\ka u_\reg \big) e^{\lambda_2(\ka u_\reg + \a)}- \big( 1 - \lambda_2\ka u_+ \big) e^{\lambda_2(\ka u_+ + \a) } \Big] .
\end{split}
\end{align}
Notice that in \eqref{I2-int}, the relevant integrals we are interested in correspond to $\lambda_1 = k_3+1$ and $\lambda_2 = k_2-k_3-1$. Substituting \eqref{u-reg-assume} into \eqref{integral-forms} and then taking the $r_0 \to 0$ limit, we get
\begin{align}\label{I2-vanish}
\begin{split}
	\lim_{r_0 \to 0} J_1^{k_3+1,k_2-k_3-1}(r_0) &= \lim_{r_0 \to 0} \frac{r_0^{\frac{1}{2}(k_2+k_3+1)}}{(k_3-k_2+1)\ka} (2\ka e^{2\ka\e})^{\frac{1}{2}(k_2-k_3-1)} \\
	&= 0 \\
	\lim_{r_0 \to 0} J_2^{k_3+1,k_2-k_3-1}(r_0) &=  \lim_{r_0 \to 0} \frac{r_0^{\frac{1}{2}(k_2+k_3+1) }}{(k_2-k_3-1)^2\ka^2} \bigg[ 1 -  \bigg(\frac{1}{2}\log(2\ka r_0) - \a \bigg) (k_2 - k_3 - 1) \bigg] \\
	&\qquad \times (2\ka e^{2\ka\e})^{\frac{1}{2}(k_2-k_3-1) }  \\
	&= 0 
\end{split}
\end{align}
where we noted in both equalities that $r_0$ is taken to a positive power, and hence each equality vanishes as $r_0 \to 0$ since all the other terms are finite or logarithmically divergent in $r_0$. It follows
\begin{align}\label{I2-fin}
\begin{split}
	\lim_{r_0 \to 0} \SI_2 = 0.
\end{split}
\end{align}

Next, the third term in the integrand in \eqref{eqn:thetastart3rneq0} is
\begin{align}
\begin{split}
	\SI_3 &\equiv \frac{\O_d r_0}{8\pi G_N} \int_{u_\reg}^{u_+} \dt u\, \p_u \Bigg(\Phi^d \frac{\d\big( \ka e^{-2(\ka u+\a)} \big)}{1-2\ka r_0 e^{-2(\ka u + \a)}} \Bigg) \\
	&= - \frac{\O_d r_0}{8\pi G_N} \Phi(u_\reg,r_0)^d \frac{(1 - 2\ka u_\reg) \d \ka - 2\ka \d\a  }{1-2\ka r_0 e^{-2(\ka u_\reg + \a)}}e^{-2(\ka u_\reg+\a)} \\
	&= - \frac{\O_d r_0}{8\pi G_N} \sum_{\substack{k_1+k_2+k_3 = d \\ k_1,k_2,k_3 \geq 0}} (-1)^{k_2+k_3} \binom{d}{k_1,k_2,k_3} L_\CB^{k_1} \bigg( \frac{1}{2\ka} e^{\ka u_\reg + \a} \bigg)^{k_2} (r_0 e^{-\ka u_\reg - \a})^{k_3}  \\
	&\qquad \times \frac{(1 - 2\ka u_\reg) \d \ka - 2\ka \d\a  }{1-2\ka r_0 e^{-2(\ka u_\reg + \a)}}e^{-2(\ka u_\reg+\a)} ,
\end{split}
\end{align}
where we used $\Phi(u_+,r_0) = 0$ and also expanded $\Phi$ via \eqref{Phi-series}. Assuming again \eqref{u-reg-assume}, we obtain\footnote{Notice now that there is a singularity in the denominator. We will see that the divergent terms cancel ultimately when evaluating \eqref{eqn:thetastart3rneq0}.}
\begin{align}
\begin{split}
	\SI_3 &= - \frac{1}{2\ka} \frac{\O_d}{8\pi G_N} \sum_{\substack{k_1+k_2+k_3 = d \\ k_1,k_2,k_3 \geq 0}} (-1)^{k_2+k_3} \binom{d}{k_1,k_2,k_3} L_\CB^{k_1} \bigg( \frac{r_0}{2\ka} \bigg)^{\frac{1}{2}(k_2+k_3)} \\
	&\qquad \times \frac{(1 - 2\ka u_\reg) \d \ka - 2\ka \d\a  }{1- e^{-2\ka\e}} e^{-2\ka\e}  .
\end{split}
\end{align}
From this it is clear if we take $r_0 \to 0$, the only terms that will contribute is when $k_2 = k_3 = 0$, which means $k_1 = d$.\footnote{Even though $u_\reg$ is divergent, we know from \eqref{u-reg-assume} that $u_\reg \sim \log r_0$. Hence, $r_0^k u_\reg \to 0$ as $r_0 \to 0$ for any $k > 0$. Notice this was used to derive \eqref{I2-vanish} above. \label{fn:ureg}} Thus, we have
\begin{align}\label{I3-fin}
\begin{split}
	\lim_{r_0 \to 0} \SI_3 &=  - \lim_{r_0 \to 0} \frac{1}{2\ka} \frac{\O_d}{8\pi G_N}  L_\CB^{d} \frac{(1 - 2\ka u_\reg) \d \ka - 2\ka \d\a  }{1- e^{-2\ka\e}} e^{-2\ka\e} .
\end{split}
\end{align}

Finally, the fourth term in the integrand in \eqref{eqn:thetastart3rneq0} is
\begin{align}
\begin{split}
	\SI_4 &\equiv \frac{\O_d r_0}{4\pi G_N} \int_{u_\reg}^{u_+} \dt u\, \Phi^d  \p_u \Bigg( \frac{\d \big( \ka e^{-2(\ka u + \a)}\big)}{1-2\ka r_0 e^{-2(\ka u + \a)}} \Bigg) \\
	&= - \frac{\O_d r_0}{4\pi G_N} \int_{u_\reg}^{u_+} \dt u\, \Phi^d \frac{4\ka e^{-2(\ka u + \a)}  }{\big( 1-2\ka r_0 e^{-2(\ka u + \a)}\big)^2}  \Big[ \big( 1 - \ka u - \ka r_0 e^{-2(\ka u + \a)} \big)\d\ka - \ka \d\a \Big] .
\end{split}
\end{align}
Again expanding $\Phi$ using \eqref{Phi-series}, we get
\begin{align}
\begin{split}
	\SI_4 &= - \frac{\O_d }{4\pi G_N} \sum_{\substack{k_1+k_2+k_3 = d \\ k_1,k_2,k_3 \geq 0} } (-1)^{k_2+k_3} \binom{d}{k_1,k_2,k_3} \frac{L_\CB^{k_1}}{(2\ka)^{k_2}}   \int_{u_\reg}^{u_+} \dt u\,   \frac{4\ka   }{\big( 1-2\ka r_0 e^{-2(\ka u + \a)}\big)^2} \\
	&\qquad\times  \Big[ ( \d\ka - \ka u \d\ka - \ka \d\a ) r_0^{k_3+1} e^{(\ka u + \a)(k_2-k_3-2)}  - \ka \d\ka r_0^{k_3+2} e^{(\ka u + \a)(k_2-k_3-4)} \Big] .
\end{split}
\end{align}
We now perform a further expansion of the denominator to get
\begin{align}\label{I4-int}
\begin{split}
	\SI_4 &= - \frac{\O_d }{4\pi G_N} \sum_{\substack{k_1+k_2+k_3 = d \\ k_1,k_2,k_3 \geq 0} } (-1)^{k_2+k_3} \binom{d}{k_1,k_2,k_3} \frac{L_\CB^{k_1}}{(2\ka)^{k_2}}  \sum_{n \geq 0}  \int_{u_\reg}^{u_+} \dt u\,   2(2\ka)^{n+1}(n+1)  \\
	&\qquad \times   \Big[ ( \d\ka - \ka u\d\ka - \ka \d\a ) r_0^{k_3+n+1} e^{(\ka u + \a)(k_2-k_3-2 - 2n)}  - \ka\d\ka  r_0^{k_3 + n +2} e^{(\ka u + \a)(k_2-k_3-4 - 2n)} \Big] \\
	&= - \frac{\O_d }{4\pi G_N} \sum_{\substack{k_1+k_2+k_3 = d \\ k_1,k_2,k_3 \geq 0} } (-1)^{k_2+k_3} \binom{d}{k_1,k_2,k_3} \frac{L_\CB^{k_1}}{(2\ka)^{k_2}}  \sum_{n \geq 0} 2(2\ka)^{n+1} (n+1) \\
	&\qquad \times   \Big[ (\d\ka -\ka\d\a) J_1^{k_3+n+1,k_2-k_3-2(n+1)} - \ka\d\ka J_2^{k_3+n+1,k_2-k_3-2(n+1)} \\
	&\qquad  - \ka\d\ka J_1^{k_3+n+2,k_2-k_3 - 2(n+2)} \Big] ,
\end{split}
\end{align}
where we rewrote the integral over $u$ in terms of the functions we defined in \eqref{integral-forms}. We are interested ultimately in the $r_0 \to 0$ limit, in which case we can ignore the terms involving $u_+$ on the right-hand side of \eqref{integral-forms} as such terms vanish. We now evaluate using \eqref{u-reg-assume}
\begin{align}\label{J123}
\begin{split}
	\lim_{r_0\to 0} J_1^{k_3+n+1,k_2-k_3-2(n+1)} &= -\lim_{r_0\to 0} \frac{(2\ka)^{\frac{1}{2}(k_2-k_3-2(n+1))} r_0^{\frac{1}{2}(k_2 + k_3)}}{(k_2-k_3-2(n+1))\ka}  e^{(k_2-k_3-2(n+1))\ka\e} \\
	\lim_{r_0 \to 0} J_2^{k_3+n+1,k_2-k_3-2(n+1)}  &= \lim_{r_0\to 0} \frac{(2\ka)^{\frac{1}{2}(k_2-k_3-2(n+1))} r_0^{\frac{1}{2}(k_2 + k_3)} }{(k_2-k_3-2(n+1))^2\ka^2}  \\
	&\qquad \times \Big[ 1- \big(k_2-k_3-2(n+1)\big) \ka u_\reg \Big] e^{(k_2-k_3-2(n+1))\ka\e} \\
	\lim_{r_0\to 0} J_1^{k_3+n+2,k_2-k_3-2(n+2)} &= - \lim_{r_0\to 0} \frac{(2\ka)^{\frac{1}{2} (k_2 - k_3 - 2(n+2)) }r_0^{\frac{1}{2}(k_2 + k_3) }}{(k_2-k_3-2(n+2))\ka}  e^{(k_2-k_3-2(n+2))\ka\e} .
\end{split}
\end{align}
Notice now for every term, only the $k_2=k_3 = 0$ term survives when we take $r_0 \to 0$, since otherwise there is a positive power of $r_0$ on the right-hand side and therefore vanishes (see Footnote~\ref{fn:ureg}). Substituting \eqref{J123} with $k_2=k_3=0$ (and hence $k_1=d$) back into \eqref{I4-int}, we get
\begin{align}\label{I4-fin}
\begin{split}
	\lim_{r_0 \to 0} \SI_4 &= - \lim_{r_0 \to 0} \frac{\O_d }{4\pi G_N} L_\CB^d \sum_{n \geq 0}   \bigg\{ \bigg[\bigg( 1 - \frac{e^{-2\ka \e}}{2} \bigg) \frac{\d\ka}{\ka} - \d\a - u_\reg \d\ka   \bigg] \\
	&\qquad   -  \frac{\d\ka}{2(n+1) \ka}   + \frac{e^{-2\ka \e}}{n+2}\frac{\d\ka}{2\ka}  \bigg\} e^{-2(n+1)\ka\e} \\
	&= - \lim_{r_0 \to 0} \frac{\O_d }{4\pi G_N} L_\CB^d \bigg[ \bigg(   \frac{2 - e^{-2\ka\e}}{2} \frac{\d\ka}{\ka} - \d\a - u_\reg\d\ka \bigg) \frac{e^{-2\ka\e}}{1-e^{-2\ka\e}}  - \frac{e^{-2\ka\e}\d\ka}{2\ka} \bigg] ,
\end{split}
\end{align}

We are now in a position to evaluate \eqref{eqn:thetastart3rneq0}, at least in the $r_0 \to 0$ limit. Using \eqref{I1-fin}, \eqref{I2-fin}, \eqref{I3-fin}, and \eqref{I4-fin}, we have
\begin{align}
\begin{split}
	\lim_{r_0 \to 0} \widetilde\Th_{\CH_s}^+ &= \lim_{r_0 \to 0} \big( \SI_1 + \SI_2 - \SI_3 + \SI_4 \big) \\
	&= - \frac{\O_d}{8\pi G_N} L_\CB^d \big( u_\reg\d\ka + \d\a \big) + \CO(\e) .
\end{split}
\end{align}
In particular, note that all the divergent terms involving $\e^{-1}$ have canceled. As this is the variation of the bulk action, we see that we can cancel this variation (recall $\d L_\CB = 0$) if we add to the bulk action a boundary action whose variation is given by
\begin{align}\label{app:reg-gn-os}
\begin{split}
	& \d I_\OS = - \lim_{\e\to0} \d \widetilde\Th^+_{\CH_s} = \frac{\O_d}{8\pi G_N} L_\CB^d \big( u_\reg \d\ka + \d\a \big)  \\
	\implies\quad & I_\OS =  \frac{A_\CB}{8\pi G_N} \big( \ka u_\reg + \a \big),
\end{split}
\end{align}
where $A_\CB \equiv \O_d L_\CB^d$ is the area of the causal diamond. This is precisely \eqref{eqn:Son-shellfixed} with $u_\CB$ replaced with $u_\mid$. Finally, the associated regularized on-shell action for the replica manifold is, as was argued in Section~\ref{subsec:replica}, obtained by taking $\ka \to \frac{\ka}{n}$, and we get
\begin{align}\label{app:reg-gn-os2}
\begin{split}
	I_\OS[n] = \frac{A_\CB}{8\pi G_N} \bigg( \frac{\ka}{n} u_\reg + \a \bigg) .
\end{split}
\end{align}

\subsection{Topological Black Hole Coordinates}\label{app:tbh-reg}

In this subsection, we will repeat the computation of the on-shell action associated to the causal diamond in topological black hole coordinates rather than Gaussian null coordinates. This provides a consistency check and also illustrates the versatility of the topological black hole coordinates. 

The topological black hole coordinates are given in the main text by \eqref{eqn:topBH}, which we reproduce here for convenience:
\begin{align}\label{top-metric}
\begin{split}
	\dt s^2 &= - 2\ka f(R) \, \dt T^2  + \frac{\dt R^2}{2\ka f(R)}   + \varphi(T,R)^{\frac{2}{d}} \, \dt\O_d^2 \\
	\varphi(T,R) &\equiv \Phi(T,R)^d, \qquad \Phi(T,R) \equiv  L - \sqrt{\frac{2 f(R)}{\ka}}\cosh ( \ka T )  , \qquad f(R) \equiv L(1+2\z) - R .
\end{split}
\end{align}
The horizon is located at
\begin{align}\label{rh}
	f(R_h) = 0 \quad\implies\quad R_h = L(1 + 2\z) .
\end{align}
Furthermore, from \eqref{map0} it is clear that constant $r$ hypersurfaces are constant $R$ hypersurfaces and hence correspond to stretched horizons. The top and bottom tips of the stretched horizon given by $R= R_0$ corresponds to $\varphi(T_\pm, R_0) = 0$. Using \eqref{top-metric}, we solve for $T_\pm$ to get
\begin{align}\label{tbh-pm}
\begin{split}
	T_\pm = \pm \frac{1}{\ka} \cosh^{-1} \Bigg( L\sqrt{\frac{\ka}{2f(R_0)} } \Bigg) .
\end{split}
\end{align}
Similarly, we can compute the $T$ coordinate of the midpoint of the stretched horizon. Noting that $\varphi$ at the midpoint is given in \eqref{phi-mid}, and we know how to relate $r$ and $R$ by \eqref{map0}, we obtain
\begin{align}\label{T-mid}
\begin{split}
	\varphi(T_\mid,R_0) = \Bigg(L - \sqrt{\frac{2f(R_0)}{\ka} }  \Bigg)^d \quad\implies\quad T_\mid = 0,
\end{split}
\end{align}
where the implication follows from the definition of $\varphi$ given in \eqref{top-metric}.

We now want to compute the pre-symplectic potential. Notice that this involves evaluating metric components on the horizon located at $R=R_h$, which may be singular. Thus, we will first evaluate on a stretched horizon $\CH_s$, defined by $R = R_0$, and take $R_0 \to R_h$ at the end. Conveniently defining
\begin{align}\label{tf-def}
	\tilde f(R) \equiv 2\ka f(R) = 2\ka L(1 + 2\z) - 2\ka R,
\end{align}
the pre-symplectic potential is after using \eqref{theta-def2-app} with $r \to R$ and evaluating the necessary Christoffel symbols
\begin{align}\label{Th-int1}
\begin{split}
	\widetilde\Theta_{\CH_s}^+ &= - \frac{1}{16\pi G_N} \int_{\CH_s^\reg} \dt T\,\dt\O_d\, \varphi(T,R_0) \Big( g^{\mu\nu}\d \G^R_{\mu\nu} - g^{\mu R} \d \G^{\l}_{\l \mu} \Big) \\
	&= \frac{\O_d}{16\pi G_N} \int_{0}^{T_+} \dt T\,  \Phi^d \bigg( \d\p_R\tf   + \frac{d \p_R\Phi}{\Phi} \d \tf + \frac{2d \tf}{\Phi} \d\p_R\Phi  \bigg)\bigg|_{R=R_0}
\end{split}
\end{align}
where as in \eqref{pre-symp-r0app} $\CH_s^\reg$ is the part of the stretched horizon that limits to $\CH^+$, with endpoints at $T=0$, which is the midpoint of $\CH_s$ by \eqref{T-mid}, and $T = T_+$, which is the top tip of $\CH_s$ by \eqref{tbh-pm}.

To further simplify \eqref{Th-int1}, note that we can use the definition of $\Phi$ given in \eqref{top-metric} and trivially express it in terms of $\tf$ defined in \eqref{tf-def} as
\begin{align}\label{Phi-app1}
	\Phi = L - \frac{\tf^{\frac{1}{2}}}{\ka}\cosh(\ka T) \quad\implies\quad \p_R\Phi =  - \frac{\p_R\tf}{2\ka \tf^{\frac{1}{2}}} \cosh(\ka T) .
\end{align}
It is simple to also evaluate their variations to be
\begin{align}\label{dPhi-1}
\begin{split}
	\d\Phi &= \d L - \frac{\d\tf}{2\ka\tf^{\frac{1}{2}}} \cosh(\ka T) - \tf^{\frac{1}{2}} \d\big( \ka^{-1}\cosh(\ka T) \big) \\
	\d\p_R\Phi &= - \frac{\d\p_R\tf}{2\ka\tf^{\frac{1}{2}}} \cosh(\ka T) + \frac{\p_R\tf\d\tf}{4\ka\tf^{\frac{3}{2}}}\cosh(\ka T) - \frac{\p_R\tf}{2\tf^{\frac{1}{2}}} \d\big( \ka^{-1}\cosh(\ka T) \big) .
\end{split}
\end{align}
Using \eqref{Phi-app1} and \eqref{dPhi-1}, it follows we can write \eqref{Th-int1} as
\begin{align}\label{Th-int2}
\begin{split}
	\widetilde\Th_{\CH_s}^+ &= \frac{\O_d}{16\pi G_N} \int_{0}^{T_+} \dt T\,  \Phi^{d-1} \bigg[ \bigg( L - \frac{(d+1)\tf^{\frac{1}{2}}}{\ka}\cosh(\ka T) \bigg) \d\p_R\tf  \\
	&\qquad  - d \tf^{\frac{1}{2}} \p_R\tf \d\big( \ka^{-1}\cosh(\ka T) \big)   \bigg]\bigg|_{R=R_0} .
\end{split}
\end{align}
In particular, note that $\d\tf$ terms have all canceled. Since $\z$ is only present in $\tf$ and not in $\p_R\tf$, we see that $\widetilde\Th^+_{\CH_s}$ does not depend on $\d\z$, and hence it does not matter whether $\z$ is a phase space constant. 

Suppose we take $R_0 \to R_h$ before performing the integral, so that we are going directly onto the causal horizon $\CH^+$. This implies $\tf \to 0$, $\p_R\tf \to - 2\ka$, $\Phi \to L$, and $T_+ \to \infty$, and so the pre-symplectic potential becomes (we drop the $\CH_s$ subscript since we are no longer on the stretched horizon)
\begin{align}\label{Th-tbh-bdy}
	\widetilde\Th^+ &= - \frac{\O_d}{8\pi G_N} \int_0^{\infty} \dt T \, L^d \d\ka \equiv  - \frac{A \d \ka}{8\pi G_N}  T_\CB, 
\end{align}
where we set $A = \O_d L^d$ and formally defined $T_\CB \to \infty$ in the last equality. This is precisely the form of the pre-symplectic potential \eqref{eqn:thetastart3} obtained in Gaussian null coordinates. To write this as a total variation, we now choose the same boundary condition as in \eqref{fixed-area}, so that we fix $A = A_\CB \equiv \O_dL_\CB^d$, the unperturbed area. It follows
\begin{align}\label{tbh-onshell1}
	\widetilde\Th^+ = -\d \bigg( \frac{A_\CB\ka T_\CB}{8\pi G_N} \bigg) \quad\implies\quad I_\OS = \frac{A_\CB}{8\pi G_N} \ka T_\CB,
\end{align}
which is precisely the unregularized on-shell action in topological black hole coordinates, and identical in form to that obtained in Gaussian null coordinates given in \eqref{eqn:Son-shellfixed}. However, as we will see below, more care is needed to evaluate the integrand on a stretched horizon since $T_+ \to \infty$ as $\tf \to 0$, and so it is not clear we can pull the limit inside the integral.

As we did in Appendix~\ref{app:gn-reg}, we will now evaluate each term separately in \eqref{Th-int2}, with the assumption both $R_0$ and $T$ are phase space constants and hence do not vary (i.e., $\d T = \d R_0 = 0$). Furthermore, as was clear in the derivation of \eqref{tbh-onshell1}, we need to fix $L = L_\CB$ as well. With these assumptions, the first term in \eqref{Th-int2} is
\begin{align}\label{J1-int}
\begin{split}
	\SJ_1 &\equiv \frac{\O_d}{16\pi G_N} \int_{0}^{T_+} \dt T\,  \Phi^{d-1} \bigg( L_\CB - \frac{(d+1)\tf(R_0)^{\frac{1}{2}}}{\ka}\cosh(\ka T) \bigg) \d\p_R\tf   \\
	&= - \frac{\O_d \d \ka }{8\pi G_N} \int_{0}^{T_+} \dt T\,  \Phi^{d-1}  \bigg( L_\CB - \frac{(d+1)\tf(R_0)^{\frac{1}{2}}}{\ka}\cosh(\ka T) \bigg) ,
\end{split}
\end{align}
where in the second equality we noted $\p_R\tf = -2\ka$. Now, for any $k \geq 1$, we perform the series expansion
\begin{align}\label{Phi-series2}
\begin{split}
	\Phi(T,R_0)^k &= \bigg( L_\CB - \frac{\tf(R_0)^{\frac{1}{2}}}{\ka}\cosh(\ka T) \bigg)^k = \sum_{\ell=0}^k \binom{k}{\ell} L_\CB^\ell(-1)^{k-\ell} \bigg( \frac{\tf(R_0)^{\frac{1}{2}}}{\ka}\cosh(\ka T) \bigg)^{k-\ell} .
\end{split}
\end{align}
It follows we can rewrite the integral in \eqref{J1-int} as a sum of integrals proportional to (keeping track of only $\tf$ and $T$ dependences)
\begin{align}\label{J-fcn-def}
\begin{split}
	J_1^{\lambda_1,\lambda_2}(R_0) &\equiv \tf(R_0)^{\lambda_1} \int_0^{T_+} \dt T\,\cosh^{\lambda_2}(\ka T) .
\end{split}
\end{align}
To evaluate this integral, we use the identity 
\begin{align}\label{useful-id}
	\cosh^{\lambda_2}(\ka T) = \frac{1}{2^{\lambda_2}}\sum_{j=0}^{\lambda_2} \binom{\lambda_2}{j} \cosh\big( (\lambda_2 - 2j) \ka T \big) .
\end{align}
Using this, we obtain
\begin{align}\label{J-fcn-fin}
\begin{split}
	J_1^{\lambda_1,\lambda_2}(R_0) &= \frac{ \tf(R_0)^{\lambda_1}}{2^{\lambda_2}} \sum_{j=0}^{\lambda_2} \binom{\lambda_2}{j}  \int_0^{T_+} \dt T\,\cosh\big( (\lambda_2 - 2j) \ka T \big) \\
	&= \frac{ \tf(R_0)^{\lambda_1}}{2^{\lambda_2}} \sum_{j=0}^{\lambda_2} \binom{\lambda_2}{j}  \frac{\sinh\big((\lambda_2-2j) \ka T_+ \big)}{(\lambda_2-2j)\ka} \\
	&= \frac{ \tf(R_0)^{\lambda_1}}{2^{\lambda_2}}  \sum_{j=0}^{\lambda_2} \Bigg( \d_{2j,\lambda_2} T_+ + (1-\d_{2j,\lambda_2}) \binom{\lambda_2}{j}  \frac{\sinh\big((\lambda_2-2j) \ka T_+ \big)}{(\lambda_2-2j)\ka} \Bigg)  ,
\end{split}
\end{align}
where in the final equality we isolated the term in the sum where $j=2\lambda_2$, in which case we have to use the fact $\frac{\sinh(n\ka T_+)}{n \ka} \to T_+$ when $n \to 0$. Writing \eqref{J1-int} in terms of the function $J_1^{\lambda_1,\lambda_2}$, we get
\begin{align}\label{J1-int2}
\begin{split}
	\SJ_1 &=  - \frac{\O_d \d \ka }{8\pi G_N} \sum_{\ell=0}^{d-1} \binom{d-1}{\ell} \frac{(-1)^{d-1-\ell}}{\ka^{d-1-\ell}} L_\CB^{\ell} \bigg[ L_\CB  J_1^{\frac{d-1-\ell}{2},d-1-\ell}(R_0)  - \frac{(d+1)}{\ka}  J_1^{\frac{d-\ell}{2}, d-\ell}(R_0) \bigg] \\
	&= - \frac{\O_d \d \ka }{8\pi G_N} \sum_{\ell=0}^{d-1} \binom{d-1}{\ell} \frac{(-1)^{d-1-\ell}}{\ka^{d-1-\ell}} L_\CB^{\ell} \Bigg[ L_\CB   \frac{ \tf^{\frac{d-1-\ell}{2}}}{2^{d-1-\ell}} \sum_{j=0}^{d-1-\ell} \Bigg( \d_{2j,d-1-\ell}T_+ \\
	&\qquad + (1 - \d_{2j,d-1-\ell} )\binom{d-1-\ell}{j}  \frac{\sinh\big((d-1-\ell-2j) \ka T_+ \big)}{(d-1-\ell-2j)\ka} \Bigg) \\
	&\qquad - \frac{(d+1)}{\ka}   \frac{ \tf^{\frac{d-\ell}{2}}}{2^{d-\ell}} \sum_{j=0}^{d-\ell} \Bigg( \d_{2j,d-\ell} T_+ + (1-\d_{2j,d-\ell})   \binom{d-\ell}{j}  \frac{\sinh\big((d-\ell-2j) \ka T_+ \big)}{(d-\ell-2j)\ka} \Bigg) \Bigg] ,
\end{split}
\end{align}
where in the second equality we used \eqref{J-fcn-fin}. Ultimately, we want to take the limit $R_0 \to R_h$, resulting in $\tf \to 0$. Therefore, any term in \eqref{J1-int2} involving a positive power of $\tf$ will vanish in this limit. Now, using \eqref{tbh-pm}, we see that $T_+ \sim \log\tf$ as $\tf \to 0$, and therefore powers of $T_+$ will not affect the previous statement. However, exponentials of $T_+$ will involve inverse powers of $\tf$ and therefore need to be treated carefully. To this end, we recall from \eqref{tbh-pm} that $T_+$ also depends on $\tf$, namely with $L = L_\CB$
\begin{align}\label{Tpm-tf}
\begin{split}
	&\cosh( \pm \ka T_\pm) = \ka L_\CB\tf^{-\frac{1}{2}}  .
\end{split}
\end{align}
As $\tf \to 0$ when $R_0 \to R_h$, to evaluate \eqref{J1-int2} in this limit, we first evaluate the series expansion
\begin{align}\label{sinh-series}
\begin{split}
	\lim_{R_0 \to R_h} \sinh\big(n \ka T_+ \big) &= \lim_{R_0 \to R_h} n\ka L_\CB \tf(R_0)^{-\frac{1}{2}}  + \CO(\tf^{\frac{1}{2}} ).
\end{split}
\end{align}
Substituting this into \eqref{J1-int2}, we get
\begin{align}\label{J1-int3}
\begin{split}
	\lim_{R_0 \to R_h} \SI_1 &= - \lim_{R_0 \to R_h} \frac{\O_d \d \ka }{8\pi G_N} \sum_{\ell=0}^{d-1} \binom{d-1}{\ell} \frac{(-1)^{d-1-\ell}}{\ka^{d-1-\ell}} L_\CB^{\ell} \Bigg[  \frac{ L_\CB}{2^{d-1-\ell}} \sum_{j=0}^{d-1-\ell} \Bigg( \d_{2j,d-1-\ell} \tf^{\frac{d-1-\ell}{2}} T_+ \\
	&\qquad + (1 - \d_{2j,d-1-\ell} )\binom{d-1-\ell}{j}  L_\CB \tf^{\frac{d-2-\ell}{2}} +  \CO(\tf^{\frac{d-\ell}{2}}) \Bigg) \\
	&\qquad - \frac{d+1}{2^{d-\ell} \ka}   \sum_{j=0}^{d-\ell} \Bigg( \d_{2j,d-\ell} \tf^{\frac{d-\ell}{2}} T_+ + (1-\d_{2j,d-\ell})   \binom{d-\ell}{j}  L_\CB \tf^{\frac{d-1-\ell}{2}} + \CO(\tf^{\frac{d+1-\ell}{2}}) \Bigg) \Bigg] .
\end{split}
\end{align}
As $\tf \to 0$ and all the other terms are finite or logarithmically divergent in $\tf$, only the terms that do not involve a positive power of $\tf$ will contribute. Noting that $\ell \leq d-1$, this means none of the higher order terms will contribute, and we arrive at
\begin{align}\label{J1-fin}
\begin{split}
	\lim_{R_0 \to R_h} \SJ_1 &=  - \lim_{R_0 \to R_h} \frac{A_\CB \d \ka }{8\pi G_N} \bigg(T_+  - \frac{2d}{\ka} \bigg) ,
\end{split}
\end{align}
where we used $A_\CB = \O_d L_\CB^d$. 

Next, the second term in the integrand in \eqref{Th-int2} is
\begin{align}\label{J2-inter}
\begin{split}
	\SJ_2 &\equiv \frac{\O_d d}{16\pi G_N}\tf(R_0)^{\frac{1}{2}} \p_R\tf \int_0^{T_+} \dt T\,\Phi^{d-1}  \d\big( \ka^{-1} \cosh(\ka T) \big) \\
	&= \frac{\O_d d \d\ka }{8\pi G_N}\tf(R_0)^{\frac{1}{2}} \int_0^{T_+} \dt T\,\Phi^{d-1}  \bigg( \frac{1}{\ka}\cosh(\ka T) - T \sinh(\ka T)  \bigg) \\
	&= \frac{\O_d d \d\ka }{8\pi G_N} \sum_{\ell=0}^{d-1} \binom{d-1}{\ell} L_\CB^\ell(-1)^{d-1-\ell} \tf(R_0)^{\frac{1}{2}} \int_0^{T_+} \dt T\, \bigg( \frac{\tf(R_0)^{\frac{1}{2}}}{\ka}\cosh(\ka T) \bigg)^{d-1-\ell}  \\
	&\qquad \times  \bigg( \frac{1}{\ka}\cosh(\ka T) - T \sinh(\ka T)  \bigg) ,
\end{split}
\end{align}
where in the second equality we expanded the variation $\d$ and also used $\p_R\tf = -2\ka$, and in the last equality we performed the series expansion \eqref{Phi-series2}. We now define the function
\begin{align}\label{J2-fcn-def}
\begin{split}
	J_2^{\lambda_1,\lambda_2}(R_0) &\equiv \tf(R_0)^{\lambda_1} \int_0^{T_+} \dt T\, T \sinh(\ka T) \cosh^{\lambda_2}(\ka T) \\
	&= \frac{\tf(R_0)^{\lambda_1}}{(\lambda_2+1)\ka} \bigg[ T\cosh^{\lambda_2+1}(\ka T) \Big|_{T=0}^{T=T_+} -  \int_0^{T_+} \dt T\, \cosh^{\lambda_2 + 1}(\ka T)  \bigg] \\
	&= \frac{1}{(\lambda_2+1)\ka}\Big( \tf(R_0)^{\lambda_1} T_+\cosh^{\lambda_2+1}(\ka T_+) - J_1^{\lambda_1,\lambda_2+1}(R_0) \Big) ,
\end{split}
\end{align}
where in the second equality we integrated by parts, and in the last equality we used the definition \eqref{J-fcn-def}. We can now write $\SJ_2$ in terms of $J_1^{\frac{d-\ell}{2},d-\ell}$ and $J_2^{\frac{d-\ell}{2},d-\ell-1}$ to obtain
\begin{align}\label{J2-int2}
\begin{split}
	\SJ_2 &= \frac{\O_d d \d\ka }{8\pi G_N} \sum_{\ell=0}^{d-1} \binom{d-1}{\ell} L_\CB^\ell \frac{(-1)^{d-1-\ell}}{\ka^{d-\ell-1}} \bigg( \frac{1}{\ka} J_1^{\frac{d-\ell}{2},d-\ell}(R_0) - J_2^{\frac{d-\ell}{2},d-\ell-1}(R_0) \bigg) \\
	&= \frac{\O_d d \d\ka }{8\pi G_N} \sum_{\ell=0}^{d-1} \binom{d-1}{\ell} L_\CB^\ell \frac{(-1)^{d-1-\ell}}{(d-\ell)\ka^{d-\ell}} \bigg( (d-\ell+1) J_1^{\frac{d-\ell}{2},d-\ell}(R_0) -  \tf^{\frac{d-\ell}{2}} T_+\cosh^{d-\ell}(\ka T_+)   \bigg) \\
	&= \frac{\O_d d \d\ka }{8\pi G_N} \sum_{\ell=0}^{d-1} \binom{d-1}{\ell} L_\CB^\ell \frac{(-1)^{d-1-\ell}}{(d-\ell)\ka^{d-\ell}} \Bigg[ \frac{ (d-\ell+1) \tf^{\frac{d-\ell}{2}}}{2^{d-\ell}} \sum_{j=0}^{d-\ell} \Bigg( \d_{2j,d-\ell} T_+ \\
	&\qquad + (1 - \d_{2j,d-\ell}) \binom{d-\ell}{j}  \frac{\sinh\big((d-\ell-2j) \ka T_+ \big)}{(d-\ell-2j)\ka} \Bigg) -  \tf^{\frac{d-\ell}{2}} T_+\cosh^{d-\ell}(\ka T_+)   \Bigg],
\end{split}
\end{align}
where in the second equality we used \eqref{J2-fcn-def}, and in the final equality we used \eqref{J-fcn-fin}. Again, we are ultimately interested in the $R_0 \to R_h$ limit, in which we will need to rewrite exponentials of $T_+$ via \eqref{Tpm-tf} and \eqref{sinh-series}. The result is
\begin{align}\label{J2-int3}
\begin{split}
	\lim_{R_0 \to R_h} \SJ_2 &= \lim_{R_0 \to R_h} \frac{\O_d d \d\ka }{8\pi G_N} \sum_{\ell=0}^{d-1} \binom{d-1}{\ell} L_\CB^\ell \frac{(-1)^{d-1-\ell}}{(d-\ell)\ka^{d-\ell}} \Bigg[ \frac{ (d-\ell+1) }{2^{d-\ell}} \sum_{j=0}^{d-\ell} \Bigg( \d_{2j,d-\ell} \tf^{\frac{d-\ell}{2}} T_+ \\
	&\qquad + (1 - \d_{2j,d-\ell}) \binom{d-\ell}{j}  L_\CB \tf^{\frac{d-1-\ell}{2}} +  \CO(\tf^{\frac{d+1-\ell}{2}})  \Bigg) -  T_+ (\ka L_\CB)^{d-\ell}  \Bigg] .
\end{split}
\end{align}
As before, when $R_0 \to R_h$, $\tf \to 0$ and all the other terms are finite or logarithmically divergent in $\tf$, so only the terms that do not involve a positive power of $\tf$ will contribute. As $\ell \leq d-1$, this means none of the higher order terms will contribute, and we arrive at
\begin{align}\label{J2-fin}
\begin{split}
	\lim_{R_0 \to R_h} \SJ_2 &=  \lim_{R_0 \to R_h} \frac{\O_d d \d\ka }{8\pi G_N}  L_\CB^{d} \Bigg(  \frac{2}{\ka}  - \sum_{\ell=0}^{d-1} \binom{d-1}{\ell} \frac{(-1)^{d-1-\ell}}{d-\ell}   T_+ \Bigg)    \\
	&= \lim_{R_0 \to R_h} \frac{A_\CB \d\ka }{8\pi G_N}  \bigg(  \frac{2d}{\ka}  -  T_+   \bigg),
\end{split}
\end{align}
where in the second equality we evaluated the sum and used $A_\CB = \O_d L_\CB^d$. 

We can now evaluate \eqref{Th-int2}, at least in the $R_0 \to 0$ limit. Using \eqref{J1-fin} and \eqref{J2-fin}, we get
\begin{align}\label{Th2-fin}
\begin{split}
	\lim_{R_0 \to R_h} \widetilde\Th^+_{\CH_s} &= \lim_{R_0 \to R_h} \big( \SJ_1 - \SJ_2 \big) = 0  .
\end{split}
\end{align}
Notice that unlike the Gaussian null case, where we recovered the same on-shell action upon regularizing (cf. \eqref{eqn:Son-shellfixed} and \eqref{app:reg-gn-os}), when we regularize the on-shell action in topological black hole coordinates we get a vanishing on-shell action! 

We now want to perform the analytic continuation done at the end of Section~\ref{subsec:replica} to get the analogous regularized Euclidean on-shell action in topological black hole coordinates. For the Gaussian null case, we analytically continued around the endpoint $(u,r) = (u_\mid, 0)$, so here we want to analytically continue around the endpoint $(T,R) = (0, R_h)$. Recalling the map between the Gaussian null coordinates and the topological black hole coordinates \eqref{map0}, we see that the analytic continuation $(u,r) \to (u,re^{-2\pi i})$ simultaneously shifts $R-R_h \to (R-R_h)e^{-2\pi i}$ and $T \to T + \frac{i\pi}{\ka}$, with the shift in $T$ necessary to keep $u$ fixed in Gaussian null coordinates. This implies that under this analytic continuation, the symplectic potential we want to evaluate is given by \eqref{Th-int2} with the endpoint $T=0$ replaced by $T = T_0 \equiv \frac{i\pi}{\ka}$, namely
\begin{align}
\begin{split}
	\lim_{R_0 \to R_h} \widetilde\Th_{\CH_s}^{+} &= \lim_{R_0 \to R_h} \frac{\O_d}{16\pi G_N} \int_{T_0}^{T_+} \dt T\,  \Phi^{d-1} \bigg[ \bigg( L - \frac{(d+1)\sqrt{\tf}}{\ka}\cosh(\ka T) \bigg) \d\p_R\tf   \\
	&\qquad - d \sqrt{\tf} \p_R\tf \d\big( \ka^{-1}\cosh(\ka T) \big)   \bigg]\bigg|_{R=R_0} \\
	&= - \lim_{R_0 \to R_h} \frac{\O_d}{16\pi G_N} \int_{0}^{T_0} \dt T\,  \Phi^{d-1} \bigg[ \bigg( L - \frac{(d+1)\sqrt{\tf}}{\ka}\cosh(\ka T) \bigg) \d\p_R\tf   \\
	&\qquad - d \sqrt{\tf} \p_R\tf \d\big( \ka^{-1}\cosh(\ka T) \big)   \bigg]\bigg|_{R=R_0},
\end{split}
\end{align}
where in the second equality we simply used the fact that the integral over the range $T \in [0,T_+]$ vanishes by \eqref{Th2-fin}, leaving us only the contribution from $0$ to $T_0$. As $T_0$ does not diverge, we can pull the limit inside the integral and set $\tf = 0$, as we did in \eqref{Th-tbh-bdy}, so that we get
\begin{align}
\begin{split}
	\lim_{R_0 \to R_h} \widetilde\Th_{\CH_s}^{+} &=  \frac{\O_d}{8\pi G_N} \int_{0}^{T_0} \dt T\,  L_\CB^{d} \d\ka = \frac{A_\CB}{8\pi G_N} T_0 \d\ka,
\end{split}
\end{align}
where we used $\p_R\tf = -2\ka$ in the first equality, and used $A_\CB = \O_dL_\CB^d$ in the final equality. The regularized on-shell action after performing the analytic continuation is then
\begin{align}
\begin{split}
	\d I_\OS = - \lim_{R_0 \to R_h} \widetilde\Th^+_{\CH_s} \quad\implies\quad I_\OS = - \frac{A_\CB\ka}{8\pi G_N} T_0 .
\end{split}
\end{align}
Substituting in $T_0 = \frac{i\pi}{\ka}$, it follows the regularized Euclidean on-shell action in topological black hole coordinates is precisely
\begin{align}
	I_\OS^\reg = -i I_\OS = - \frac{A_\CB}{8G_N} ,
\end{align}
which matches \eqref{onshell-reg-n} with $n=1$ obtained from Gaussian null coordinates.

\section{Photon Phase Shift From Shapiro Time Delay}\label{app:photon}

In this appendix, we demonstrate that we can also obtain the total phase shift obtained in \eqref{phase-shift-fin} by considering the Shapiro time delay, which is given in \cite{Lee:2024oxo} to be
\begin{align}\label{eqn:Shapiro}
\begin{split}
	\D\tau_{\text{Shapiro}} = - \int_{-L}^0 \dt\tt\, n_{\mu}^+ \d \tilde U^\mu_+ - \int_0^L \dt\tt \, n^-_\mu \d \tilde U^\mu_-, 
\end{split}
\end{align}
where 
\begin{align}
\begin{split}
	n^{\pm\mu} = \p_\tt \pm \p_\tr , \qquad n^\pm_\mu = - \dt\tt \pm \dt\tr
\end{split}
\end{align}
is the tangent vector (and tangent one-form) of $\CH^\pm$, and
\begin{align}\label{delta-U}
\begin{split}
	\d \tilde U_\pm^\mu &\equiv \frac{\dt x^{\mu}}{\dt \tt} - n^{\pm \mu}
\end{split}
\end{align}
is the difference between the velocity vector along $\CH_s$ and that along $\CH^\pm$, each normalized to have norm $-1$ rather than $-m_\eff^2$. Along $\CH_s$, we have
\begin{align}\label{Hs-vec}
\begin{split}
	\frac{\dt x^\mu}{\dt\tt} = \frac{1}{m_\eff} U^\mu \frac{\dt\tau}{\dt \tt} =  \p_\tt - \frac{\tt}{\sqrt{\tt^2 + s_\CB^2}} \p_\tr ,
\end{split}
\end{align}
where in the first equality we used the definition of the four-velocity given in \eqref{4-velocity} and divided by $m_\eff$ due to the different normalizations used, and in the last equality we used \eqref{eqn:u} and also \eqref{eqn:effmass0} to set $\lambda = m_\eff$. Substituting this into \eqref{delta-U} and then into \eqref{eqn:Shapiro}, we find
\begin{align}\label{shapiro-final}
\begin{split}
	\D\tau_{\text{Shapiro}} &=  2 L + \int_{-L}^0 \dt\tt  \frac{\tt}{\sqrt{\tt^2 + s_\CB^2}} - \int_0^L \dt\tt  \frac{\tt}{\sqrt{\tt^2 + s_\CB^2}} \\
	&= 2s_\CB + \CO(s_\CB^2).
\end{split}
\end{align}
Multiplying by the laser frequency $\w_0$, we get precisely the total phase shift \eqref{phase-shift-fin}. 

Finally, we return to the simple example given in Section~\ref{subsec:fluc-horizon}, where we demonstrated how causal horizons in one coordinate system parametrizes a stretched horizon in another coordinate system. Clearly, care should be taken when directly applying a linearized analysis such as that of \cite{Lee:2024oxo} to metric perturbations of the form \eqref{eqn:hexpl} when $\tilde{t} \sim s_{\CB}$. Nevertheless, using the full metric \eqref{eqn:flatmetric}, without expanding in small $s_{\CB}$, the proper time observable along the beamsplitter is given by
\begin{equation}
	2 \int_0^L \dt\tilde{t} \, \frac{\tilde{t}}{\sqrt{\tilde{t}^2+s_{\CB}^2}} = 2\Big(\sqrt{L^2 + s_\CB^2} - s_\CB \Big)  .
\end{equation}
It follows the shift in proper time along the origin is given by
\begin{align}
\begin{split}
	\D \tau = 2 L - 2\Big(\sqrt{L^2 + s_\CB^2} - s_\CB \Big)   = 2s_\CB + \CO(s_\CB^2),
\end{split}
\end{align}
which is precisely the Shapiro time delay \eqref{shapiro-final}.

\bibliography{rindler_fluc.bib}{}

\providecommand{\href}[2]{#2}\begingroup\raggedright\begin{thebibliography}{10}

\bibitem{Bekenstein:1973ur}
J.~D. Bekenstein, ``{Black holes and entropy},''
  \href{http://dx.doi.org/10.1103/PhysRevD.7.2333}{{\em Phys. Rev. D}
  {\bfseries 7} (1973) 2333--2346}.

\bibitem{Hawking:1975vcx}
S.~W. Hawking, ``{Particle Creation by Black Holes},''
  \href{http://dx.doi.org/10.1007/BF02345020}{{\em Commun. Math. Phys.}
  {\bfseries 43} (1975) 199--220}. [Erratum: Commun.Math.Phys. 46, 206 (1976)].

\bibitem{Gibbons:1976ue}
G.~W. Gibbons and S.~W. Hawking, ``{Action Integrals and Partition Functions in
  Quantum Gravity},'' \href{http://dx.doi.org/10.1103/PhysRevD.15.2752}{{\em
  Phys. Rev. D} {\bfseries 15} (1977) 2752--2756}.

\bibitem{Witten:2024upt}
E.~Witten, ``{Introduction to Black Hole Thermodynamics},''
  \href{http://arxiv.org/abs/2412.16795}{{\ttfamily arXiv:2412.16795
  [hep-th]}}.

\bibitem{Bekenstein:1975tw}
J.~D. Bekenstein, ``{Statistical Black Hole Thermodynamics},''
  \href{http://dx.doi.org/10.1103/PhysRevD.12.3077}{{\em Phys. Rev. D}
  {\bfseries 12} (1975) 3077--3085}.

\bibitem{Zurek:1985gd}
W.~H. Zurek and K.~S. Thorne, ``{Statistical mechanical origin of the entropy
  of a rotating, cha rged black hole},''
  \href{http://dx.doi.org/10.1103/PhysRevLett.54.2171}{{\em Phys. Rev. Lett.}
  {\bfseries 54} (1985) 2171}.

\bibitem{Ryu:2006bv}
S.~Ryu and T.~Takayanagi, ``{Holographic derivation of entanglement entropy
  from AdS/CFT},'' \href{http://dx.doi.org/10.1103/PhysRevLett.96.181602}{{\em
  Phys. Rev. Lett.} {\bfseries 96} (2006) 181602},
  \href{http://arxiv.org/abs/hep-th/0603001}{{\ttfamily arXiv:hep-th/0603001}}.

\bibitem{Ryu:2006ef}
S.~Ryu and T.~Takayanagi, ``{Aspects of Holographic Entanglement Entropy},''
  \href{http://dx.doi.org/10.1088/1126-6708/2006/08/045}{{\em JHEP} {\bfseries
  08} (2006) 045}, \href{http://arxiv.org/abs/hep-th/0605073}{{\ttfamily
  arXiv:hep-th/0605073}}.

\bibitem{Hayden:2007cs}
P.~Hayden and J.~Preskill, ``{Black holes as mirrors: Quantum information in
  random subsystems},''
  \href{http://dx.doi.org/10.1088/1126-6708/2007/09/120}{{\em JHEP} {\bfseries
  09} (2007) 120}, \href{http://arxiv.org/abs/0708.4025}{{\ttfamily
  arXiv:0708.4025 [hep-th]}}.

\bibitem{Almheiri:2012rt}
A.~Almheiri, D.~Marolf, J.~Polchinski, and J.~Sully, ``{Black Holes:
  Complementarity or Firewalls?},''
  \href{http://dx.doi.org/10.1007/JHEP02(2013)062}{{\em JHEP} {\bfseries 02}
  (2013) 062}, \href{http://arxiv.org/abs/1207.3123}{{\ttfamily arXiv:1207.3123
  [hep-th]}}.

\bibitem{Harlow:2014yka}
D.~Harlow, ``{Jerusalem Lectures on Black Holes and Quantum Information},''
  \href{http://dx.doi.org/10.1103/RevModPhys.88.015002}{{\em Rev. Mod. Phys.}
  {\bfseries 88} (2016) 015002},
  \href{http://arxiv.org/abs/1409.1231}{{\ttfamily arXiv:1409.1231 [hep-th]}}.

\bibitem{Penington:2019npb}
G.~Penington, ``{Entanglement Wedge Reconstruction and the Information
  Paradox},'' \href{http://dx.doi.org/10.1007/JHEP09(2020)002}{{\em JHEP}
  {\bfseries 09} (2020) 002}, \href{http://arxiv.org/abs/1905.08255}{{\ttfamily
  arXiv:1905.08255 [hep-th]}}.

\bibitem{Laflamme:1987ec}
R.~Laflamme, ``{Entropy of a Rindler Wedge},''
  \href{http://dx.doi.org/10.1016/0370-2693(87)90799-4}{{\em Phys. Lett. B}
  {\bfseries 196} (1987) 449--450}.

\bibitem{Casini:2011kv}
H.~Casini, M.~Huerta, and R.~C. Myers, ``{Towards a derivation of holographic
  entanglement entropy},''
  \href{http://dx.doi.org/10.1007/JHEP05(2011)036}{{\em JHEP} {\bfseries 05}
  (2011) 036}, \href{http://arxiv.org/abs/1102.0440}{{\ttfamily arXiv:1102.0440
  [hep-th]}}.

\bibitem{Hung:2011nu}
L.-Y. Hung, R.~C. Myers, M.~Smolkin, and A.~Yale, ``{Holographic Calculations
  of Renyi Entropy},'' \href{http://dx.doi.org/10.1007/JHEP12(2011)047}{{\em
  JHEP} {\bfseries 12} (2011) 047},
  \href{http://arxiv.org/abs/1110.1084}{{\ttfamily arXiv:1110.1084 [hep-th]}}.

\bibitem{Perlmutter:2013gua}
E.~Perlmutter, ``{A universal feature of CFT R{\'e}nyi entropy},''
  \href{http://dx.doi.org/10.1007/JHEP03(2014)117}{{\em JHEP} {\bfseries 03}
  (2014) 117}, \href{http://arxiv.org/abs/1308.1083}{{\ttfamily arXiv:1308.1083
  [hep-th]}}.

\bibitem{Jacobson:2015hqa}
T.~Jacobson, ``{Entanglement Equilibrium and the Einstein Equation},''
  \href{http://dx.doi.org/10.1103/PhysRevLett.116.201101}{{\em Phys. Rev.
  Lett.} {\bfseries 116} no.~20, (2016) 201101},
  \href{http://arxiv.org/abs/1505.04753}{{\ttfamily arXiv:1505.04753 [gr-qc]}}.

\bibitem{Verlinde:2019ade}
E.~Verlinde and K.~M. Zurek, ``{Spacetime Fluctuations in AdS/CFT},''
  \href{http://dx.doi.org/10.1007/JHEP04(2020)209}{{\em JHEP} {\bfseries 04}
  (2020) 209}, \href{http://arxiv.org/abs/1911.02018}{{\ttfamily
  arXiv:1911.02018 [hep-th]}}.

\bibitem{Jacobson:2022gmo}
T.~Jacobson and M.~R. Visser, ``{Entropy of causal diamond ensembles},''
  \href{http://dx.doi.org/10.21468/SciPostPhys.15.1.023}{{\em SciPost Phys.}
  {\bfseries 15} no.~1, (2023) 023},
  \href{http://arxiv.org/abs/2212.10608}{{\ttfamily arXiv:2212.10608
  [hep-th]}}.

\bibitem{Ciambelli:2019lap}
L.~Ciambelli, R.~G. Leigh, C.~Marteau, and P.~M. Petropoulos, ``{Carroll
  Structures, Null Geometry and Conformal Isometries},''
  \href{http://dx.doi.org/10.1103/PhysRevD.100.046010}{{\em Phys. Rev. D}
  {\bfseries 100} no.~4, (2019) 046010},
  \href{http://arxiv.org/abs/1905.02221}{{\ttfamily arXiv:1905.02221
  [hep-th]}}.

\bibitem{Chandrasekaran:2021hxc}
V.~Chandrasekaran, E.~E. Flanagan, I.~Shehzad, and A.~J. Speranza,
  ``{Brown-York charges at null boundaries},''
  \href{http://dx.doi.org/10.1007/JHEP01(2022)029}{{\em JHEP} {\bfseries 01}
  (2022) 029}, \href{http://arxiv.org/abs/2109.11567}{{\ttfamily
  arXiv:2109.11567 [hep-th]}}.

\bibitem{Freidel:2022vjq}
L.~Freidel and P.~Jai-akson, ``{Carrollian hydrodynamics and symplectic
  structure on stretched horizons},''
  \href{http://dx.doi.org/10.1007/JHEP05(2024)135}{{\em JHEP} {\bfseries 05}
  (2024) 135}, \href{http://arxiv.org/abs/2211.06415}{{\ttfamily
  arXiv:2211.06415 [gr-qc]}}.

\bibitem{Ciambelli:2023mir}
L.~Ciambelli, L.~Freidel, and R.~G. Leigh, ``{Null Raychaudhuri: canonical
  structure and the dressing time},''
  \href{http://dx.doi.org/10.1007/JHEP01(2024)166}{{\em JHEP} {\bfseries 01}
  (2024) 166}, \href{http://arxiv.org/abs/2309.03932}{{\ttfamily
  arXiv:2309.03932 [hep-th]}}.

\bibitem{Ciambelli:2024swv}
L.~Ciambelli, L.~Freidel, and R.~G. Leigh, ``{Quantum null geometry and
  gravity},'' \href{http://dx.doi.org/10.1007/JHEP12(2024)028}{{\em JHEP}
  {\bfseries 12} (2024) 028}, \href{http://arxiv.org/abs/2407.11132}{{\ttfamily
  arXiv:2407.11132 [hep-th]}}.

\bibitem{Chandrasekaran:2023vzb}
V.~Chandrasekaran and E.~E. Flanagan, ``{Horizon phase spaces in general
  relativity},'' \href{http://dx.doi.org/10.1007/JHEP07(2024)017}{{\em JHEP}
  {\bfseries 07} (2024) 017}, \href{http://arxiv.org/abs/2309.03871}{{\ttfamily
  arXiv:2309.03871 [hep-th]}}.

\bibitem{Bub:2024nan}
M.~W. Bub, T.~He, P.~Mitra, Y.~Zhang, and K.~M. Zurek, ``{The Quantum Mechanics
  of a Spherically Symmetric Causal Diamond in Minkowski Spacetime},''
  \href{http://arxiv.org/abs/2408.11094}{{\ttfamily arXiv:2408.11094
  [hep-th]}}.

\bibitem{Bisognano:1975ih}
J.~J. Bisognano and E.~H. Wichmann, ``{On the Duality Condition for a Hermitian
  Scalar Field},'' \href{http://dx.doi.org/10.1063/1.522605}{{\em J. Math.
  Phys.} {\bfseries 16} (1975) 985--1007}.

\bibitem{Bisognano:1976za}
J.~J. Bisognano and E.~H. Wichmann, ``{On the Duality Condition for Quantum
  Fields},'' \href{http://dx.doi.org/10.1063/1.522898}{{\em J. Math. Phys.}
  {\bfseries 17} (1976) 303--321}.

\bibitem{Jacobson:2018ahi}
T.~Jacobson and M.~Visser, ``{Gravitational Thermodynamics of Causal Diamonds
  in (A)dS},'' \href{http://dx.doi.org/10.21468/SciPostPhys.7.6.079}{{\em
  SciPost Phys.} {\bfseries 7} no.~6, (2019) 079},
  \href{http://arxiv.org/abs/1812.01596}{{\ttfamily arXiv:1812.01596
  [hep-th]}}.

\bibitem{Jacobson:2022jir}
T.~Jacobson and M.~R. Visser, ``{Partition Function for a Volume of Space},''
  \href{http://dx.doi.org/10.1103/PhysRevLett.130.221501}{{\em Phys. Rev.
  Lett.} {\bfseries 130} no.~22, (2023) 221501},
  \href{http://arxiv.org/abs/2212.10607}{{\ttfamily arXiv:2212.10607
  [hep-th]}}.

\bibitem{Harlow:2018tqv}
D.~Harlow and D.~Jafferis, ``{The Factorization Problem in Jackiw-Teitelboim
  Gravity},'' \href{http://dx.doi.org/10.1007/JHEP02(2020)177}{{\em JHEP}
  {\bfseries 02} (2020) 177}, \href{http://arxiv.org/abs/1804.01081}{{\ttfamily
  arXiv:1804.01081 [hep-th]}}.

\bibitem{Balasubramanian:2025hns}
V.~Balasubramanian and T.~Yildirim, ``{How to Count States in Gravity},''
  \href{http://arxiv.org/abs/2506.15767}{{\ttfamily arXiv:2506.15767
  [hep-th]}}.

\bibitem{Callan:1994py}
C.~G. Callan, Jr. and F.~Wilczek, ``{On geometric entropy},''
  \href{http://dx.doi.org/10.1016/0370-2693(94)91007-3}{{\em Phys. Lett. B}
  {\bfseries 333} (1994) 55--61},
  \href{http://arxiv.org/abs/hep-th/9401072}{{\ttfamily arXiv:hep-th/9401072}}.

\bibitem{Holzhey:1994we}
C.~Holzhey, F.~Larsen, and F.~Wilczek, ``{Geometric and renormalized entropy in
  conformal field theory},''
  \href{http://dx.doi.org/10.1016/0550-3213(94)90402-2}{{\em Nucl. Phys. B}
  {\bfseries 424} (1994) 443--467},
  \href{http://arxiv.org/abs/hep-th/9403108}{{\ttfamily arXiv:hep-th/9403108}}.

\bibitem{Calabrese:2004eu}
P.~Calabrese and J.~L. Cardy, ``{Entanglement entropy and quantum field
  theory},'' \href{http://dx.doi.org/10.1088/1742-5468/2004/06/P06002}{{\em J.
  Stat. Mech.} {\bfseries 0406} (2004) P06002},
  \href{http://arxiv.org/abs/hep-th/0405152}{{\ttfamily arXiv:hep-th/0405152}}.

\bibitem{Lewkowycz:2013nqa}
A.~Lewkowycz and J.~Maldacena, ``{Generalized gravitational entropy},''
  \href{http://dx.doi.org/10.1007/JHEP08(2013)090}{{\em JHEP} {\bfseries 08}
  (2013) 090}, \href{http://arxiv.org/abs/1304.4926}{{\ttfamily arXiv:1304.4926
  [hep-th]}}.

\bibitem{Shekar:2024edg}
A.~Shekar and M.~Taylor, ``{Replica analysis of entanglement properties},''
  \href{http://dx.doi.org/10.1103/PhysRevD.111.066018}{{\em Phys. Rev. D}
  {\bfseries 111} no.~6, (2025) 066018},
  \href{http://arxiv.org/abs/2410.07312}{{\ttfamily arXiv:2410.07312
  [hep-th]}}.

\bibitem{DeBoer:2018kvc}
J.~De~Boer, J.~J\"arvel\"a, and E.~Keski-Vakkuri, ``{Aspects of capacity of
  entanglement},'' \href{http://dx.doi.org/10.1103/PhysRevD.99.066012}{{\em
  Phys. Rev. D} {\bfseries 99} no.~6, (2019) 066012},
  \href{http://arxiv.org/abs/1807.07357}{{\ttfamily arXiv:1807.07357
  [hep-th]}}.

\bibitem{Banks:2021jwj}
T.~Banks and K.~M. Zurek, ``{Conformal description of near-horizon vacuum
  states},'' \href{http://dx.doi.org/10.1103/PhysRevD.104.126026}{{\em Phys.
  Rev. D} {\bfseries 104} no.~12, (2021) 126026},
  \href{http://arxiv.org/abs/2108.04806}{{\ttfamily arXiv:2108.04806
  [hep-th]}}.

\bibitem{Verlinde:2022hhs}
E.~Verlinde and K.~M. Zurek, ``{Modular fluctuations from shockwave
  geometries},'' \href{http://dx.doi.org/10.1103/PhysRevD.106.106011}{{\em
  Phys. Rev. D} {\bfseries 106} no.~10, (2022) 106011},
  \href{http://arxiv.org/abs/2208.01059}{{\ttfamily arXiv:2208.01059
  [hep-th]}}.

\bibitem{Gukov:2022oed}
S.~Gukov, V.~S.~H. Lee, and K.~M. Zurek, ``{Near-horizon quantum dynamics of 4D
  Einstein gravity from 2D Jackiw-Teitelboim gravity},''
  \href{http://dx.doi.org/10.1103/PhysRevD.107.016004}{{\em Phys. Rev. D}
  {\bfseries 107} no.~1, (2023) 016004},
  \href{http://arxiv.org/abs/2205.02233}{{\ttfamily arXiv:2205.02233
  [hep-th]}}.

\bibitem{He:2024vlp}
T.~He, A.-M. Raclariu, and K.~M. Zurek, ``{An Infrared On-Shell Action and its
  Implications for Soft Charge Fluctuations in Asymptotically Flat
  Spacetimes},'' \href{http://arxiv.org/abs/2408.01485}{{\ttfamily
  arXiv:2408.01485 [hep-th]}}.

\bibitem{Ciambelli:2025flo}
L.~Ciambelli, T.~He, and K.~M. Zurek, ``{Quantum Area Fluctuations from
  Gravitational Phase Space},''
  \href{http://arxiv.org/abs/2504.12282}{{\ttfamily arXiv:2504.12282
  [hep-th]}}.

\bibitem{Aalsma:2025bcg}
L.~Aalsma and S.-E. Bak, ``{Modular Fluctuations in Cosmology},''
  \href{http://arxiv.org/abs/2503.04886}{{\ttfamily arXiv:2503.04886
  [hep-th]}}.

\bibitem{Verlinde:2019xfb}
E.~P. Verlinde and K.~M. Zurek, ``{Observational signatures of quantum gravity
  in interferometers},''
  \href{http://dx.doi.org/10.1016/j.physletb.2021.136663}{{\em Phys. Lett. B}
  {\bfseries 822} (2021) 136663},
  \href{http://arxiv.org/abs/1902.08207}{{\ttfamily arXiv:1902.08207 [gr-qc]}}.

\bibitem{Damour:1979wya}
T.~Damour, {\em {Quelques proprietes mecaniques, electromagnetiques,
  thermodynamiques et quantiques des trous noir}}.
\newblock PhD thesis, Paris U., VI-VII, 1979.

\bibitem{Price:1986yy}
R.~H. Price and K.~S. Thorne, ``{Membrane Viewpoint on Black Holes: Properties
  and Evolution of the Stretched Horizon},''
  \href{http://dx.doi.org/10.1103/PhysRevD.33.915}{{\em Phys. Rev. D}
  {\bfseries 33} (1986) 915--941}.

\bibitem{Thorne:1986iy}
K.~S. Thorne, R.~H. Price, and D.~A. Macdonald, eds., {\em {BLACK HOLES: THE
  MEMBRANE PARADIGM}}.
\newblock 1986.

\bibitem{Susskind:1993if}
L.~Susskind, L.~Thorlacius, and J.~Uglum, ``{The Stretched horizon and black
  hole complementarity},''
  \href{http://dx.doi.org/10.1103/PhysRevD.48.3743}{{\em Phys. Rev. D}
  {\bfseries 48} (1993) 3743--3761},
  \href{http://arxiv.org/abs/hep-th/9306069}{{\ttfamily arXiv:hep-th/9306069}}.

\bibitem{Parikh:1997ma}
M.~Parikh and F.~Wilczek, ``{An Action for black hole membranes},''
  \href{http://dx.doi.org/10.1103/PhysRevD.58.064011}{{\em Phys. Rev. D}
  {\bfseries 58} (1998) 064011},
  \href{http://arxiv.org/abs/gr-qc/9712077}{{\ttfamily arXiv:gr-qc/9712077}}.

\bibitem{Carroll:2004st}
S.~M. Carroll, \href{http://dx.doi.org/10.1017/9781108770385}{{\em {Spacetime
  and Geometry}: {An Introduction to General Relativity}}}.
\newblock Cambridge University Press, 7, 2019.

\bibitem{Dong:2013qoa}
X.~Dong, ``{Holographic Entanglement Entropy for General Higher Derivative
  Gravity},'' \href{http://dx.doi.org/10.1007/JHEP01(2014)044}{{\em JHEP}
  {\bfseries 01} (2014) 044}, \href{http://arxiv.org/abs/1310.5713}{{\ttfamily
  arXiv:1310.5713 [hep-th]}}.

\bibitem{Dong:2016fnf}
X.~Dong, ``{The Gravity Dual of Renyi Entropy},''
  \href{http://dx.doi.org/10.1038/ncomms12472}{{\em Nature Commun.} {\bfseries
  7} (2016) 12472}, \href{http://arxiv.org/abs/1601.06788}{{\ttfamily
  arXiv:1601.06788 [hep-th]}}.

\bibitem{Nakaguchi:2016zqi}
Y.~Nakaguchi and T.~Nishioka, ``{A holographic proof of R{\'e}nyi entropic
  inequalities},'' \href{http://dx.doi.org/10.1007/JHEP12(2016)129}{{\em JHEP}
  {\bfseries 12} (2016) 129}, \href{http://arxiv.org/abs/1606.08443}{{\ttfamily
  arXiv:1606.08443 [hep-th]}}.

\bibitem{Colin-Ellerin:2020mva}
S.~Colin-Ellerin, X.~Dong, D.~Marolf, M.~Rangamani, and Z.~Wang, ``{Real-time
  gravitational replicas: Formalism and a variational principle},''
  \href{http://dx.doi.org/10.1007/JHEP05(2021)117}{{\em JHEP} {\bfseries 05}
  (2021) 117}, \href{http://arxiv.org/abs/2012.00828}{{\ttfamily
  arXiv:2012.00828 [hep-th]}}.

\bibitem{Colin-Ellerin:2021jev}
S.~Colin-Ellerin, X.~Dong, D.~Marolf, M.~Rangamani, and Z.~Wang, ``{Real-time
  gravitational replicas: low dimensional examples},''
  \href{http://dx.doi.org/10.1007/JHEP08(2021)171}{{\em JHEP} {\bfseries 08}
  (2021) 171}, \href{http://arxiv.org/abs/2105.07002}{{\ttfamily
  arXiv:2105.07002 [hep-th]}}.

\bibitem{Banihashemi:2024weu}
B.~Banihashemi and T.~Jacobson, ``{The enigmatic gravitational partition
  function},'' \href{http://dx.doi.org/10.1007/s10714-024-03347-0}{{\em Gen.
  Rel. Grav.} {\bfseries 57} no.~2, (2025) 43},
  \href{http://arxiv.org/abs/2411.00267}{{\ttfamily arXiv:2411.00267
  [hep-th]}}.

\bibitem{Glorioso:2018mmw}
P.~Glorioso, M.~Crossley, and H.~Liu, ``{A prescription for holographic
  Schwinger-Keldysh contour in non-equilibrium systems},''
  \href{http://arxiv.org/abs/1812.08785}{{\ttfamily arXiv:1812.08785
  [hep-th]}}.

\bibitem{Schwinger:1960qe}
J.~S. Schwinger, ``{Brownian motion of a quantum oscillator},''
  \href{http://dx.doi.org/10.1063/1.1703727}{{\em J. Math. Phys.} {\bfseries 2}
  (1961) 407--432}.

\bibitem{Keldysh:1964ud}
L.~V. Keldysh, ``{Diagram technique for nonequilibrium processes},''
  \href{http://dx.doi.org/10.1142/9789811279461_0007}{{\em Zh. Eksp. Teor.
  Fiz.} {\bfseries 47} (1964) 1515--1527}.

\bibitem{Bousso:1999xy}
R.~Bousso, ``{A Covariant entropy conjecture},''
  \href{http://dx.doi.org/10.1088/1126-6708/1999/07/004}{{\em JHEP} {\bfseries
  07} (1999) 004}, \href{http://arxiv.org/abs/hep-th/9905177}{{\ttfamily
  arXiv:hep-th/9905177}}.

\bibitem{Balasubramanian:2013rqa}
V.~Balasubramanian, B.~Czech, B.~D. Chowdhury, and J.~de~Boer, ``{The entropy
  of a hole in spacetime},''
  \href{http://dx.doi.org/10.1007/JHEP10(2013)220}{{\em JHEP} {\bfseries 10}
  (2013) 220}, \href{http://arxiv.org/abs/1305.0856}{{\ttfamily arXiv:1305.0856
  [hep-th]}}.

\bibitem{Banks:2020tox}
T.~Banks, P.~Draper, and S.~Farkas, ``{Path Integrals for Causal Diamonds and
  the Covariant Entropy Principle},''
  \href{http://dx.doi.org/10.1103/PhysRevD.103.106022}{{\em Phys. Rev. D}
  {\bfseries 103} no.~10, (2021) 106022},
  \href{http://arxiv.org/abs/2008.03449}{{\ttfamily arXiv:2008.03449
  [hep-th]}}.

\bibitem{Lee:2024oxo}
V.~S.~H. Lee and K.~M. Zurek, ``{Proper time observables of general
  gravitational perturbations in laser interferometry-based gravitational wave
  detectors},'' \href{http://dx.doi.org/10.1103/6q7d-jz26}{{\em Phys. Rev. D}
  {\bfseries 111} no.~12, (2025) 124037},
  \href{http://arxiv.org/abs/2408.03363}{{\ttfamily arXiv:2408.03363
  [hep-ph]}}.

\bibitem{Ebadi:2023gne}
R.~Ebadi, D.~E. Kaplan, S.~Rajendran, and R.~L. Walsworth, ``{GALILEO: Galactic
  Axion Laser Interferometer Leveraging Electro-Optics},''
  \href{http://dx.doi.org/10.1103/PhysRevLett.132.101001}{{\em Phys. Rev.
  Lett.} {\bfseries 132} no.~10, (2024) 101001},
  \href{http://arxiv.org/abs/2306.02168}{{\ttfamily arXiv:2306.02168
  [hep-ph]}}.

\bibitem{Zhang:2023mkf}
Y.~Zhang and K.~M. Zurek, ``{Stochastic description of near-horizon
  fluctuations in Rindler-AdS},''
  \href{http://dx.doi.org/10.1103/PhysRevD.108.066002}{{\em Phys. Rev. D}
  {\bfseries 108} no.~6, (2023) 066002},
  \href{http://arxiv.org/abs/2304.12349}{{\ttfamily arXiv:2304.12349
  [hep-th]}}.

\bibitem{Bak:2024kzk}
S.-E. Bak, C.~Keeler, Y.~Zhang, and K.~M. Zurek, ``{Rindler fluids from
  gravitational shockwaves},''
  \href{http://dx.doi.org/10.1007/JHEP05(2024)331}{{\em JHEP} {\bfseries 05}
  (2024) 331}, \href{http://arxiv.org/abs/2403.18013}{{\ttfamily
  arXiv:2403.18013 [hep-th]}}.

\bibitem{He:2014laa}
T.~He, V.~Lysov, P.~Mitra, and A.~Strominger, ``{BMS supertranslations and
  Weinberg{\textquoteright}s soft graviton theorem},''
  \href{http://dx.doi.org/10.1007/JHEP05(2015)151}{{\em JHEP} {\bfseries 05}
  (2015) 151}, \href{http://arxiv.org/abs/1401.7026}{{\ttfamily arXiv:1401.7026
  [hep-th]}}.

\bibitem{He:2020ifr}
T.~He and P.~Mitra, ``{Covariant Phase Space and Soft Factorization in
  Non-Abelian Gauge Theories},''
  \href{http://dx.doi.org/10.1007/JHEP03(2021)015}{{\em JHEP} {\bfseries 03}
  (2021) 015}, \href{http://arxiv.org/abs/2009.14334}{{\ttfamily
  arXiv:2009.14334 [hep-th]}}.

\bibitem{Kapec:2022hih}
D.~Kapec, ``{Soft particles and infinite-dimensional geometry},''
  \href{http://dx.doi.org/10.1088/1361-6382/ad0514}{{\em Class. Quant. Grav.}
  {\bfseries 41} no.~1, (2024) 015001},
  \href{http://arxiv.org/abs/2210.00606}{{\ttfamily arXiv:2210.00606
  [hep-th]}}.

\bibitem{He:2023bvv}
T.~He and P.~Mitra, ``{Asymptotic structure of higher dimensional Yang-Mills
  theory},'' \href{http://dx.doi.org/10.21468/SciPostPhys.16.5.142}{{\em
  SciPost Phys.} {\bfseries 16} no.~5, (2024) 142},
  \href{http://arxiv.org/abs/2306.04571}{{\ttfamily arXiv:2306.04571
  [hep-th]}}.

\bibitem{Caminiti:2025hjq}
J.~Caminiti, F.~Capeccia, L.~Ciambelli, and R.~C. Myers, ``{Geometric modular
  flows in 2d CFT and beyond},''
  \href{http://arxiv.org/abs/2502.02633}{{\ttfamily arXiv:2502.02633
  [hep-th]}}.

\end{thebibliography}\endgroup
\bibliographystyle{utphys}
	
\end{document}